\documentclass[superscriptaddress,groupedaddress,nofootnoteinbib,11pt]{article}
\pdfoutput=1 
\usepackage{jheppub}

\usepackage{amsmath, amsfonts, amsthm, amssymb, graphicx, color,hyperref}
\usepackage{subfigure}

\usepackage[utf8]{inputenc}

\allowdisplaybreaks[3]

\def \bal#1\eal  {\begin{align} #1 \end{align}}
\newcommand{\be} {\begin{equation}}
\newcommand{\ee} {\end{equation}}
\newcommand{\nn} {\nonumber\\}

\newcommand{\ud} {\mathrm{d}}

\newcommand{\dd} {\delta}

\newcommand{\pd} {\partial}
\newcommand{\bfx} {{\bf x}}
\newcommand{\bfy} {{\bf y}}

\newcommand{\mc} {\mathcal}

\newcommand{\mn} {{\mu\nu}}

\newcommand{\ai}{{\alpha}}
\newcommand{\bi}{{\beta}}

\newcommand{\ri}{{\rho}}
\newcommand{\si}{{\sigma}}
\newcommand{\li}{{\lambda}}

\newcommand{\ep}{{\epsilon}}

\newcommand{\para}[1]{\par\vspace{2mm}\noindent\textbf{{#1}}.---}
\def\ba{\begin{eqnarray}}
\def\ea{\end{eqnarray}}

\def\L{\mathcal{L}}

\def\stu{St\"uckelberg }

\def\d{\mathrm{d}}
\def\mn{_{\mu \nu}}
\def\mupn{^\mu_{\, \nu}}
\def\({\left(}
\def\){\right)}

\def\mpl{M_{\rm Pl}}
\def\p{\partial}
\def\ie{{\em i.e. }}
\def\ien{{\em i.e.}}

\title{The $\Lambda_2$ limit of massive gravity}
\author{Claudia de Rham, Andrew J. Tolley and Shuang-Yong Zhou}
\affiliation{CERCA, Department of Physics, Case Western Reserve University, 10900 Euclid Ave, Cleveland, OH 44106, USA}
\date{\today}

\abstract{
Lorentz--invariant massive gravity is usually associated with a strong coupling scale $\Lambda_3$. 
By including non--trivial effects from the \stu modes, we show that about these vacua, one can push the strong coupling scale to higher values and evade the linear  vDVZ--discontinuity. For generic parameters of the theory and generic vacua for the \stu fields, the $\Lambda_2$--decoupling limit of the theory is well--behaved and free of any ghost or gradient--like instabilities. We also discuss the implications for nonlinear sigma models with Lorentzian target spaces.
}

\begin{document}
\maketitle
\flushbottom

\section{Introduction and Summary}

As an effective field theory on Minkowski space, Lorentz--invariant massive gravity with generic interactions is strongly coupled and breaks perturbative unitarity at a scale $\Lambda_*$ with $\Lambda_*< \Lambda_3= (\mpl m^{2})^{{1/3}}$~\cite{ArkaniHamed:2002sp}. When the graviton mass $m$ is taken to be of the current Hubble scale, this is a very small scale phenomenologically. Moreover, all the interactions that arise strictly below the scale  $\Lambda_3$ are associated with the nonlinear Boulware--Deser (BD) ghost~\cite{Boulware:1973my,Deffayet:2005ys,Creminelli:2005qk}. This makes the Vainshtein mechanism~\cite{Vainshtein:1972sx} in all these massive gravity theories untrustworthy as a resolution of the linear vDVZ--discontinuity (van Dam--Veltman--Zakharov~\cite{vanDam:1970vg,Zakharov:1970cc}). As a result, none of the theories of massive gravity with a strong coupling scale $\Lambda_*< \Lambda_3$ have a smooth massless limit to General Relativity  within the regime of validity of their  effective field theory.\\

Fortunately, all the interactions below $\Lambda_3$ can be eliminated by a unique graviton potential~\cite{deRham:2010ik,deRham:2010kj}, and this coincides with the elimination of the BD ghost~\cite{deRham:2010kj,Hassan:2011hr,Hassan:2011ea}. In ghost--free massive gravity~\cite{deRham:2010ik,deRham:2010kj} gravitational waves carry 5 modes, as expected for a massive spin--2  particle in four dimensions, and the Vainshtein mechanism operates in a much more controlled way~\cite{Deffayet:2001uk}.
See~\cite{deRham:2014zqa} for a recent review of massive gravity and~\cite{Babichev:2013usa} for an introduction on the Vainshtein mechanism.

\subsection*{$\Lambda_2$--limit of massive gravity}
The scale $\Lambda_3=(\mpl m^{2})^{{1/3}}$ is usually considered as the highest possible strong coupling scale in a Lorentz--invariant theory of massive gravity (bearing in mind we consider $m\lll \mpl$). This usually comes from analyzing ghost--free massive gravity around the trivial Lorentz--invariant vacuum $g_{\mu\nu}=\eta_{\mu\nu},~\phi^A = x^A$, where the $\phi^A$ are the \stu scalar fields that ensure that the theory of massive gravity is diffeomorphism invariant.

However, in ghost--free massive gravity (also known as the dRGT model~\cite{deRham:2010ik,deRham:2010kj}\;\footnote{Since we are mainly interested in Lorentz--invariant massive gravity, the reference metric is chosen as the Minkowski metric~\cite{deRham:2010ik,deRham:2010kj} in most parts of the paper. Analogous analysis and conclusion apply to non-Minkowski references metrics. In particular, in \S \ref{sec:massgra}, when discussing non-compact nonlinear sigma models, we consider non-Minkowski reference (target) metrics.}), about non--trivial vacua
\be
\label{nontrivial vac}
g_{\mu\nu}=\eta_{\mu\nu} +\mc{O}(m^2),~~~ \phi^A  = \bar\phi^A(x) \ne x^A\,,
\ee
which still preserves approximate Lorentz--invariance for the metric (in the limit where $m\to 0$) but not for the \stu fields, the associated strong coupling scale can be parametrically higher than $\Lambda_3$.
In unitary gauge, the metric for the non--trivial vacuum configuration~\eqref{nontrivial vac} is still approximately Minkowski (and hence Lorentz--invariant) but in a different coordinate form,  $g_{\mu\nu}=\pd_\mu\tilde\phi^A \pd_\nu\tilde\phi^B \eta_{AB}+\mc{O}(m^2)$, with $\tilde\phi^A(x)$ being the inverse function of $\bar\phi^A(x)$.\\

We will show this in a couple of different ways. First of all, writing the metric as $g\mn=\eta\mn+h\mn/\mpl$,
we note that if all the vector and scalar modes obtain a kinetic term without needing to rely on a mixing with $h\mn$, then one can define a $\Lambda_2$--decoupling limit for ghost--free massive gravity ({\it i.e.}, the dRGT model),  by sending
\ba
\label{eq:limit}
\mpl \to \infty, ~~~~ m \to 0, ~~~~ \Lambda_2=\sqrt{\mpl m} \to {\rm fixed}  ,
\ea
which leads to
\ba
\label{introdl}
S_{\rm GFMG} \to  \int\ud^4 x \( - \frac{1}{4}{h}^{\mu\nu} \mc{E}^{\ri\si}_{\mu\nu} {h}_{\ri\si}  +  \Lambda_2^4   \L_{\rm MG-NLS}[\phi^A] +\frac{h\mn}{2\mpl} T^{\mu\nu}\)\,,
\ea
where $S_{\rm GFMG}$ is given by Eq.~(\ref{mastergravitylag}), the first term is the linear Einstein--Hilbert term, $T^{\mu\nu}$ is the stress--energy tensor of the matter fields and  we have defined the {\it massive gravity nonlinear sigma model} as
\ba
\label{MG-NLSint}
\L_{\rm MG-NLS}[\phi^A]=\sum_{n=2}^{4} \ai_n {K}^{\mu_1}_{[\mu_1}  {K}^{\mu_2}_{\mu_2} \cdots {K}^{\mu_n}_{\mu_n]}\,,
\ea
with $K^\mu_\nu=\dd^\mu_\nu - X^\mu_\nu$ with ${X}^\mu_\nu = \sqrt{\eta^{\mu\ri}\pd_\ri \phi^A \pd_\nu \phi^B \eta_{AB}}$.
The interesting properties of this nonlinear sigma model and its generalization have been discussed in~\cite{deRham:2015ijs} and will also be mentioned later in this paper.\\

We emphasize that the massive gravity nonlinear sigma model (\ref{MG-NLSint}) does  not amount to simply setting $g_{\mu\nu}:=\eta_{\mu\nu}$ in ghost--free massive gravity, which would be an  inconsistent procedure. Rather, we take a well--defined $\Lambda_2$--decoupling limit which preserves the total number of degrees of freedom along the flow $\mpl \rightarrow \infty$, and hence will automatically carry over desirable properties of ghost--free massive gravity (such as the absence of the BD ghost) to the decoupled theory.
This fact alone is sufficient to guarantee that $ \L_{\rm MG-NLS}[\phi^A] $ does not carry more that 3 propagating degrees of freedom (in $D=4$ dimensions), while the full action \eqref{introdl} still carries all the 5 propagating degrees of freedom.
The very existence of such a decoupling limit relies on configurations for $\phi^A$ for which all 3 propagating degrees of freedom in $\L_{\rm MG-NLS}[\phi^A] $ are active. \\

In what follows we will first perform a full nonlinear Hamiltonian analysis for this massive gravity nonlinear sigma model. That is, we run a Dirac--Bergmann algorithm for the model, finding out all the constraints and checking their consistencies. We stress again that since we are taking a consistent decoupling limit, it is guaranteed that the number of degrees of freedom is not more than three, since $h_{\mu\nu}$ accounts for the additional two.
For technical reasons, we will limit ourselves to the so--called minimal model although our results hold in all generality for generic sets of parameters. As expected, this Hamiltonian analysis concludes that in four dimensions,  $3$ out of the $4$ \stu fields are dynamical degrees of freedom. In other words, both the vector and scalar modes in $\phi^A$ are dynamical. Interestingly, even though `gravity' is entirely decoupled, the BD ghost mode is still eliminated. This of course is due to the matrix square root structure and the anti--symmetization scheme of the ghost--free graviton potential~\cite{deRham:2011rn} and was guaranteed by taking the decoupling limit. \\

Having proven that the nonlinear sigma model~\eqref{MG-NLSint} includes 3 degrees of freedom one can then search for backgrounds where the longitudinal mode is dynamical. In principle most vacua of the theory will excite all 3 DoFs, but the trivial one $\langle \phi^A\rangle=x^A$ and any Lorentz--invariant generalization are special in that at linear order they exhibit an accidental $U(1)$--gauge symmetry. For the isolated nonlinear sigma model, the longitudinal mode is thus infinitely strongly coupled on these trivial vacua and their regime of validity is null. For massive gravity, however, the coupling to gravity breaks the accidental $U(1)$ and provides a kinetic term for all the relevant degrees of freedom. This implies that vacua where the \stu fields preserve Lorentz--invariance are acceptable vacua for massive gravity and the strong coupling scale on these vacua is lowered to $\Lambda_3$, but these vacua are not acceptable for the nonlinear sigma model.
Instead, for the nonlinear sigma model and for massive gravity with a $\Lambda_2$--decoupling limit, one needs to consider non--trivial (weakly Lorentz--breaking) vacua for the \stu fields. (Of course, for the nonlinear sigma model alone, $\Lambda_2$ is a free tunable dimensionfull parameter.) \\

Finding exact vacua may be generically challenging from a purely technical viewpoint. Plane waves are exact solutions which play the role of instructive toy--models. More generic vacua can be constructed perturbatively, either by performing a small field expansion about the trivial vacuum or by performing a local expansion about a given point in spacetime.
The latter expansion will prove convenient to establish the full stability of the DoFs and derive the corresponding strong coupling scale.\\

A nontrivial background $\bar\phi^a$ will necessarily introduce some characteristic energy scale $L^{-1}$ (it may of course introduce more  scales and when that happens, the relevant energy scale for this discussion is the smallest one). When taking the decoupling limit~\eqref{eq:limit} we maintain the scale $L^{-1}$ fixed
 and the resulting strong coupling scale ends up being $\Lambda_2$ dressed by some positive powers of $L^{-1}$. 
This scale $L^{-1}$ plays a similar role as the anti--de Sitter (AdS) curvature  when considering massive gravity on AdS~\cite{Kogan:2000uy,Porrati:2000cp,Karch:2001jb,Porrati:2003sa,Porrati:2004mz}. Note however that unlike massive gravity on AdS, we will focus this discussion to the case where the spacetime curvature vanishes (at least up to order $m^2$ corrections).

\subsection*{Absence of linear vDVZ--discontinuity\footnote{We emphasize that for most theories of massive gravity, the vDVZ discontinuity is arising from considering a linear theory beyond its regime of validity, and represents a failure of the linear theory; while  the discontinuity is expected to be absent at the non--linear level.
In this manuscript we show that in a large class of non--trivial vacua, the {\it absence of the discontinuity} is already manifest at the {\it linear level} for ghost--free massive gravity.}}

The previous $\Lambda_2$--decoupling limit of ghost--free massive gravity has another virtue: Namely the absence of coupling between matter fields and the \stu fields. Indeed in the decoupled limit~\eqref{introdl}, only the standard tensor modes $h\mn$ couple to matter as in General Relativity while the additional three degrees of freedom and specifically the longitudinal mode fully decouple. This immediately implies that already in the linear regime, \ie already at large distances compared to $ L $ and $\Lambda_2^{-1}$ but smaller than $m^{-1}$, the phenomenology of ghost--free massive gravity on these vacua is very close to General Relativity, without even needing to invoke any explicit Vainshtein mechanism (or in other words the non--trivial vacua already automatically implement the Vainshtein mechanism). Beyond this decoupling limit we expect corrections suppressed by positive powers of $\Lambda_*/\mpl$, and fifth forces will also be suppressed by a similar amount (see Ref.~\cite{deRham:2012ew} for relevant discussions). \\

The decoupling of the longitudinal mode  also implies that the theory ({\it i.e.}, ghost-free massive gravity {\it with the Minkowski reference metric}) is free from the standard vDVZ--discontinuity at the linearized level about these non--trivial vacua, similarly as for massive gravity on AdS~\cite{Kogan:2000uy,Porrati:2000cp,Karch:2001jb} (or a general FLRW background~\cite{deRham:2010tw,Aoki:2015xqa}). A crucial distinction with massive gravity on AdS is that in our approach  the gravitational (or geometric) sector is insensitive to the scale $ L $ in the decoupling limit and the background metric is Minkowski--like (or can be taken to be de Sitter or FLRW if the relevant cosmological constant or matter fields are included). For massive gravity on AdS on the other hand, the gravitational sector is strongly sensitive to the AdS curvature scale $ L $ even in the decoupling limit. For massive gravity on AdS, setting a limit where the metric is Minkowski requires sending $ L ^{-1}\to 0$ and therefore leads to an arbitrarily low strong coupling scale (see Fig.~\ref{Fig:Limits}).\\

Our approach also differs from standard Lorentz--violating theories of massive gravity (see Ref.~\cite{Comelli:2014xga} for a classification), where the strong coupling scale can be $\Lambda_2$ (or even higher when considering Lorentz--breaking generalizations of the Einstein--Hilbert term~\cite{DeFelice:2015hla}). Indeed in these theories, the Lagrangian {\it manifestly} breaks Lorentz invariance. In the model we consider here, the fundamental theory preserves Lorentz invariance and the latter is only broken {\it spontaneously} about the vacua we consider.

\subsection*{Nonlinear sigma models with Lorentzian target spaces}

The potential of massive gravity can be seen as a non--standard nonlinear sigma model for the four \stu fields $\phi^A$,
mapping from the spacetime metric $g\mn$ (or $\eta\mn$ in the absence of gravity) to the target space (the reference metric~\cite{deRham:2010gu}).\\

For a standard nonlinear sigma model, a typical requirement is that the target space be Riemannian (its metric being positive definite) to avoid ghost DoFs (see e.g.~\cite{nsmrev,nsmrev2,Zakrzewski:1989na}). From this point of view, it is not surprising that generically massive gravity is plagued by the BD ghost, as the internal space of the \stu fields is Lorentzian (pseudo--Riemannian with signature $(-+\cdots+)$). Ghost--free massive gravity then acts as a unique and special case that evades the Riemannian requirement. For a symmetric target space, the Lorentzian nature translates to non--compactness of the associated symmetry group.\\

At the technical level, the reason why the BD ghost is eliminated in ghost--free massive gravity is due to the existence of a second--class constraint~\cite{deRham:2010ik}. Taking the decoupling limit~\eqref{eq:limit} of ghost--free massive gravity, the nonlinear sigma model decouples from the gravitational tensor DoFs. Since a decoupling limit never changes the number of DoFs (if taken appropriately\footnote{All the examples where a decoupling limit seemingly changes the number of DoFs are arising from improperly taking the decoupling limit. See \S~8.1 of~\cite{deRham:2014zqa} for examples and a discussion on this point.}), the absence of the sixth BD mode in ghost--free massive gravity ensures the absence of ghost in the nonlinear  sigma model.
As a result and as we mentioned above, the nonlinear sigma model that arises from massive gravity is free of the ghost associated with the negative direction of the target space. This is in contrast with the other known ways to avoid the Riemannian requirement of the target space which all rely on invoking some gauge DoFs. This is for instance the case of the string Polyakov/Nambu--Goto action~\cite{Brink:1976sc,Deser:1976rb, Polyakov:1975rr, Nambu:1986ze, Goto:1971ce,Hara:1971ur}, or more generally for  $p$--brane actions~\cite{Becker:2007zj}, where the target space is the spacetime itself, thus Lorentzian. Another known mechanism is to invoke normal gauge fields that are auxiliary, that is, without a kinetic term for the gauge field. This mechanism is used in supergravity model building~(see e.g.~\cite{Cremmer:1979up,VanNieuwenhuizen:1981ae}). All these known exceptions with a Lorentzian target space do not compromise the spirit of the Riemannian requirement in the sense that once the auxiliary gauge/diffeomorphism DoFs  are fixed by making use of the auxiliary field equations of motion and gauge choices the target space becomes manifestly Riemannian. On the other hand, the massive gravity nonlinear sigma model and its generalization relies on two second class constraints to project out the would--be ghost associated with the negative direction.\\

Since the ghost--free graviton potential is unique, up to a few free parameters, it follows that the massive gravity nonlinear sigma model~\eqref{MG-NLSint} in $D$ dimensions -- with the sum starting from $n=1$, the internal space metric $\eta_{AB}$ replaced by $f_{AB}(\phi)$ and the coefficients $\ai_n$ generalized to be functions of the \stu fields $\ai_n(\phi)$~\cite{deRham:2014lqa,deRham:2015ijs} -- is the only nonlinear sigma model where the target space is Lorentzian. We emphasize that the target space can be higher--dimensional than that of the spacetime (that is $N>D$). The case of $N<D$ is more subtle and will be discussed in \S~\ref{sec:sum}. See~\cite{deRham:2015ijs} for a bi--gravity braneworld interpretation of this generalized nonlinear sigma model and more discussions on nonlinear sigma models with Lorentzian target spaces.\\

\noindent {\bf Outline.--} The rest of the manuscript is organized as follows: We start by introducing ghost--free massive gravity and a generalization of the Nambu--Goto action in \S~\ref{sec:massgra}, derive the value of the strong coupling scale about the trivial vacuum on Minkowski and AdS, and explain the origin of the vDVZ--discontinuity on Minkowski and its absence on AdS. We then perform the full nonlinear Hamiltonian analysis in \S~\ref{sec:nha} for the massive gravity nonlinear sigma model and confirm the existence of two second class constraints that remove the BD ghost associated with the negative direction of the target space.
Motivated by this result we first provide in \S~\ref{sec:NonTrivialVacua} an explicit exact nonlinear example of vacuum solution  where all the DoFs are manifest. Although that vacuum turns out to be unstable, it corresponds to a useful explicit proof--of--principle. In \S~\ref{sec:gpb} we then derive more general classes of backgrounds by expanding the background itself and by adopting a local coordinate expansion. We find a family of stable vacua where all the DoFs are manifest and healthy. The related strong coupling scale on these stable vacua is established in \S~\ref{sec:strong}. These results are valid in dimensions larger than two. In two--dimensions we show in \S~\ref{sec:2d} that the $U(1)$--symmetry is preserved to all orders and the corresponding nonlinear sigma model hence propagates no DoFs.
In \S~\ref{sec:sum}, we give a short summary of our main results.\\

\noindent {\bf Conventions.--}  In what follows we use a convention where Greek letters denote $D$--dimensional spacetime indices while capital latin letters $A,B,\ldots$ denote indices in the $N\geq D$ Lorentzian internal space metric. When $N=D$ we may also use Greek letters to designate the internal space indices. Lowercase latin letters $i,j,\ldots$ designate spatial indices. In $D$--dimensions we use the convention for the strong coupling scale $\Lambda_r=(m^{r-1}\mpl ^{\frac{D-2}{2}})^{\frac{2}{D+2r-4}}$.


\section{Ghost--free massive gravity and nonlinear sigma model}
\label{sec:massgra}

In this section, we introduce the ghost--free graviton potential in a conceptually novel way: As a non--standard nonlinear sigma model with a Lorentzian target space. In this formulation, the importance of the scale $\Lambda_2$ is manifest.

\subsection{Nambu--Goto action for  non--compact space}

We start by considering a theory of $N$ scalar fields $\phi^A$ living on a $D$--dimensional flat spacetime metric $\eta\mn$. These $N$ scalar fields  may be thought as coordinates of a non--trivial target (field space, or internal) manifold specified by the metric $f_{AB}(\phi)$. This corresponds to a  nonlinear sigma model whose action can typically be written as
\ba
\label{Sigma1}
\L_{\Sigma} =   -\frac12\eta^{\mu\nu} \pd_\mu \phi^A \pd_\nu \phi^B f_{AB}(\phi)  -V(\phi)\,.
\ea
Nonlinear sigma models~\cite{GellMann:1960np} are effective field theories for multiple fields $\phi^A$ with applications in various areas of physics (see, e.g.,~\cite{nsmrev,nsmrev2,Zakrzewski:1989na} and references therein for a review). The nonlinear sigma model of Eq.~(\ref{Sigma1}) is well--defined and free of ghost if the internal space metric  $f_{AB}$ is positive definite, \ien, the target space has to be Riemannian (as opposed to pseudo--Riemannian). If the target space is symmetric, this means that the associated isometry group needs to be compact.\\

When considering a non--compact space, the internal space metric $f_{AB}$ typically has a negative eigenvalue and the sigma model~\eqref{Sigma1} has a ghost. One possible way out is to ensure that the mode associated with the negative direction is in fact not dynamical or a gauge mode. This is indeed the resolution for the Polyakov action for a $p$--brane\,\footnote{The Polyakov and Nambu--Goto action usually refer to the one dimensional string actions, but they have a simple generalization to a $p$--brane on a $p+1$ dimensional world volume.} where the spacetime metric $\eta\mn$ is promoted to an auxiliary field $g\mn(x)$ and diffeomorphism invariance ensures that the would--be ghost DoF associated with the negative direction of the internal space is a gauge mode:
\be
\L_{\rm Polyakov} = - \frac{\sqrt{-g}}2    \bigg[ g^{\mu\nu} \pd_\mu \phi^{A}(x) \pd_\nu \phi^{B}(x) f_{AB}(\phi)  -  p + 1 \bigg]  \,.
\ee
 If the internal space has signature $(-+\cdots+)$, then naively the field $\phi^0(x)$ behaves as a ghost. But this action is invariant under the diffeomorphisms and the naive ghost is merely a gauge degree of freedom. This is obvious in the `static' gauge where $\phi^\mu=x^\mu$, for  $\mu=0,\ldots,p$, and the left--over target space is manifestly positive definite for the remaining $\phi^A$ with $A=p+1,\ldots,N-1$.
An alternative way to see this is to write the auxiliary field metric $g\mn$ in ADM form \cite{Arnowitt:1962hi} $g\mn \d x^\mu \d x^\nu=-\(N^0\)^2 \d t^2+\gamma_{ij}\(\d x^i+N^i \d t\)\(\d x^j+N^j \d t\)$, and then the lapse $N^0$ plays the role of a Lagrange multiplier that imposes a first class constraint projecting out the would--be ghost DoF,
 \ba
 \mathcal{H}_{\rm Polyakov}\supset \frac 12 N^0\(\frac{1}{\sqrt{\gamma}}f^{AB}p_A p_B+\sqrt{\gamma} \( \gamma^{ij}\p_i\phi^A\p_j\phi^B f_{AB} -p+1 \) \)\,,
 \ea
 with $p_A=\p \L_{\rm Polyakov} / \p \dot \phi^A$, and we have accounted for the entire dependence on the lapse in the Hamiltonian.  Actually, for the $p=1$ string case, we see that for this procedure to work it is essential that the internal space metric $f_{AB}$ be not sign definite, otherwise the constraint would fix more than one phase space variable. In addition to this Hamiltonian constraint, there are $D-1$ additional first class constraints generated by the shifts $N^i$ but only the Hamiltonian constraint is required to remove the would--be ghost in this Lorentzian space.\\

Since the metric $g\mn$ is not dynamical in this model and merely plays the role of auxiliary variables, we can integrate it out without changing the number of DoFs, and we are then left with the well--known Nambu--Goto action for the $p$--brane:
\be
\label{NGaction}
\L_{\rm NG} =  \sqrt{-\det(\pd_\mu \phi^{A} \pd_\nu \phi^{B} f_{AB}(\phi))}\,.
\ee
The Nambu--Goto action still enjoys the  same gauge symmetry, and static gauge can still be chosen to make the target space manifestly positive definite. \\

On the other hand, if the $D$--dimensional tensor  $X\mupn$ defined as\footnote{The matrix square root is taken as the principal branch solution of the matrix equation $X^\mu_{\, \alpha} X^\alpha_{\, \nu} = g^{\mu\alpha} \pd_\alpha \phi^A \pd_\nu \phi^B f_{AB}$. }
\ba
\label{eq:X}
X\mupn =X\mupn[f_{AB}, \phi^A] = \sqrt{\eta^{\mu\alpha}\pd_\alpha \phi^A \pd_\nu \phi^B f_{AB}}
\ea
is diagonalizable, then the Nambu--Goto action may also be re--written as
\ba
\label{NGeta}
\L_{\rm NG}=\det X= X^{\mu_1}_{\, [\mu_1}X^{\mu_2}_{\, \mu_2}\cdots X^{\mu_D}_{\, \mu_D]}   ,
\ea
where our anti--symmetrization convention is with the averaging factor $1/n!$ in front.
In this language, the absence of ghost for this non--compact target space can be traced back to
\ba
\label{eq:detHessian1}
\det \(\frac{\p^2 \L_{\rm NG}}{\p \dot \phi^A \p \dot \phi^B}\)=0\,,
\ea
signaling that not all of the $N$ scalar fields $\phi^A$ are dynamical.\\

\subsection{Generalization of Nambu--Goto}

Inspired by the expression (\ref{NGeta}) for the Nambu--Goto action, it is now natural to extend it to the following Lagrangians for $n\le D$,
\ba
\label{LNt}
\tilde \L_n=X^{\mu_1}_{\, [\mu_1} X^{\mu_2}_{\, \mu_2}\cdots X^{\mu_n}_{\, \mu_n]}\,,
\ea
so that $\tilde \L_D\equiv \L_{\rm NG}$ and $\tilde \L_0\equiv1$. We may also consider a fully equivalent representation of the $\tilde \L_n$ by taking linear combinations of them and defining the following Lagrangians
\ba
\label{LN}
\L_n=K^{\mu_1}_{\, [\mu_1} K^{\mu_2}_{\, \mu_2}\cdots K^{\mu_n}_{\, \mu_n]}\,,
\ea
with $K\mupn=\delta\mupn-X\mupn$. So long as $N\ge D$,  all of these Lagrangians for any $0\le n \le D$ satisfy the same relation~\eqref{eq:detHessian1} as the Nambu--Goto action, namely,
\ba
\det \(\frac{\p^2 \L_{n}}{\p \dot \phi^A \p \dot \phi^B}\)=0\,,
\ea
which ensures the absence of ghost in any of these theories. While this generalization seems to be natural mathematically or at a superficial level, there is a crucial difference between the Nambu--Goto action and the generalized Lagrangians considered in~\eqref{LN}: For the Nambu--Goto action, the rank of the matrix $\mathcal{H}_{AB}={\p^2 \L_{n}}/{\p \dot \phi^A \p \dot \phi^B}$ is $N-D$, while for the $\L_n$ the rank of the associated matrix $\mathcal{H}_{AB}$ is $N-1$. Also as we have seen, the removal of the degrees of freedom for the Nambu--Goto action is associated with a gauge symmetry, while for the other $\L_n$ no symmetry is present and the removal of the ghost is related to second--class constraints. Nevertheless, for each one of these Lagrangians the vanishing of the Hessian is what signals the absence of the would--be ghost for these $\Sigma$--models on the target space. Therefore, the generalized Nambu--Goto Lagrangian (which we will refer to as the massive gravity nonlinear sigma model for reasons to become clear shortly) is given by
\ba
\label{MG-NLS}
\L_{\rm MG-NLS}=\sum_{n=1}^{D} \ai_n {K}^{\mu_1}_{[\mu_1}  {K}^{\mu_2}_{\mu_2} \cdots {K}^{\mu_n}_{\mu_n]}\,,
\ea
where
\be
K\mupn =\dd^\mu_\nu - \sqrt{\eta^{\mu\alpha}\pd_\alpha \phi^A \pd_\nu \phi^B f_{AB}(\phi)},
\ee
and  $N\ge D$. \\

Intriguingly, this generalization of the $p$--brane Nambu--Goto action exactly gives rise to the graviton potential of ghost--free massive gravity when $N=D$. To consider in the context of a curved spacetime, we note that, instead of Eq.~(\ref{NGeta}), the Nambu--Goto action can equivalently be casted as
\ba
\L_{\rm NG}=\sqrt{-g}\det \mc{X}=\sqrt{-g} \mc{X}^{\mu_1}_{\, [\mu_1}\mc{X}^{\mu_2}_{\, \mu_2}\cdots \mc{X}^{\mu_D}_{\, \mu_D]} ,~~  {\rm with}~\mc{X}\mupn = \sqrt{g^{\mu\ri}\pd_\ri \phi^A \pd_\nu \phi^B f_{AB}(\phi)}\,.
\ea
The generalization of this action to terms with fewer factors  of $\mc{X}$ is exactly the ghost--free graviton potential. The difference again is that, while the Nambu--Goto term is diffeomorphism invariant, the terms with fewer factors of $\mc{X}$ are not. \\

In what follows we will also consider embedding these models in a gravitational setup, \ien, coupling to the dynamical part of $g_{\mu\nu}$. This leads to ghost--free massive gravity in $D$ dimensions (we shall consider only $N=D$ in the following, as our main interest is in the context of massive gravity)~\cite{deRham:2010ik,deRham:2010kj}
\ba
\label{mastergravitylag}
S_{\rm GFMG} &=&\mpl^{D-2} \int \ud^{D} x \sqrt{-g} \left( \frac{R}{2} -  \Lambda_c +  m^2\sum_{n=2}^{D} \ai_n \mc{K}^{\mu_1}_{[\mu_1}  \mc{K}^{\mu_2}_{\mu_2} \cdots \mc{K}^{\mu_n}_{\mu_n]}  \right) \\
&=&\mpl^{D-2} \int \ud^{D} x \sqrt{-g} \left( \frac{R}{2} -  \Lambda'_c +  m^2\sum_{n=1}^{D-1} \bi_n \mc{X}^{\mu_1}_{[\mu_1}  \mc{X}^{\mu_2}_{\mu_2} \cdots \mc{X}^{\mu_n}_{\mu_n]}  \right)   ,
\ea
where
\ba
\label{eq:KandX}
\mc{K}\mupn \equiv \delta^\mu_\nu - \mc{X}^\mu_\nu,~~~ {\rm with}~~\mc{X}\mupn = \sqrt{g^{\mu\ri}\pd_\ri \phi^A \pd_\nu \phi^B f_{AB}(\phi)}\,,
\ea
and the fields $\phi^A$ play the role of \stu fields that restore diffeomorphism invariance. In the gravitational setup, $\L_1$ is a tadpole term and $ \mc{X}^{\mu_1}_{[\mu_1}  \mc{X}^{\mu_2}_{\mu_2} \cdots \mc{X}^{\mu_D}_{\mu_D]}$ acts as a cosmological constant so we do not consider their contributions. Without loss of generality we may always set $\alpha_2=1$ and $\bi_1=-1$. The constants $\ai_n$ and $\bi_n$ are related via
\be
\alpha_i = (-1)^i \sum_{j=i}^{D}  C_{D-i}^{D-j} \beta_j,~~~~ \beta_i = (-1)^i \sum_{j=i}^{D}  C_{D-i}^{D-j} \alpha_j,
 ~~~{\rm with}~~ C_n^m\equiv \frac{n!}{m!(n-m)!}   ,
\ee
where $\ai_1\equiv 0,~\ai_2\equiv 1,\bi_1\equiv -1, ~\bi_D\equiv 0$. $\ai_{2<n<D-1}$ and $\bi_{1<n<D}$ are two sets of equivalent free parameters of ghost--free massive gravity.\\

\subsection{Linearized theory on Minkowski}
\label{sec:l3limit}

\para{``$\Sigma$--model"}
Before considering the effects of gravity, we first focus on the ``potential" term of massive gravity as a Lagrangian for the scalar fields $\phi^A$ in their own right living on a flat Minkowski spacetime, decoupled from the gravitational sector. Note that {\it a priori} it is not certain that this ``potential'' scalar theory from massive gravity is actually continuously connected to massive gravity, which would require the existence of a decoupling limit of some sort.
We will see that such a decoupling limit indeed exists. At any rate, for now, one may consider the ``potential'' action of massive gravity as a scalar field theory on its own. Let us, for instance, consider the following Lagrangian
\ba
\label{eq:L2}
\L_2[\phi^A,\eta_{AB}]=K^\mu_{\, [\nu} K^\nu_{\mu]}\,.
\ea
As will be shown in section~\ref{sec:nha}, non--perturbatively, this Lagrangian carries $D-1$ degrees of freedom (the constraint that removes the ghost in ghost--free massive gravity remains active even in the absence of gravity). However,  perturbatively about the trivial vacuum $\phi^A=x^\mu \delta^A_\mu$, the Lagrangian~\eqref{eq:L2} only carries $D-2$ rather than $D-1$ DoFs. Indeed, at the linearized level, $\phi^A=x^\mu \delta^A_\mu+V^A$, the Lagrangian $\L_2$ is a Maxwell theory for $V^A$ and enjoys a $U(1)$ gauge symmetry. In dimensions $N=D> 2$, that symmetry is an artifact of the linearized theory and does not survive at the nonlinear level.\\

This realization has a profound impact not only for the scalar theory~\eqref{eq:L2}, but also for massive gravity as we shall see later. Indeed, for the scalar theory~\eqref{eq:L2}, the fact that one DoF fails to be dynamical on the trivial vacuum $\phi^a=x^a$ implies that this vacuum is infinitively strongly coupled and cannot be trusted (its has no regime of validity). This means that the theory~\eqref{eq:L2} only makes sense if considered about different non--trivial vacua which excites all $D-1$ degrees of freedom.\\

\para{Implications for massive gravity}
In the context of massive gravity the situation is more positive for the vacuum $\phi^A=x^A$. Indeed the mixing with gravity breaks the $U(1)$ gauge symmetry and all $D-1$ DoFs in the fields $\phi^A$ are dynamical. The trivial vacuum $\phi^\alpha=x^\alpha$ has then an interesting non--trivial regime of validity.
In this case one of the DoF in $\phi^\alpha$ only becomes dynamical (at the linearized level) through its mixing  with gravity. This implies that, at the linear level, this DoF directly couples to matter with the same strength as gravity, which is at the origin of the linear vDVZ discontinuity.\\

To see this explicitly, let us start with the ghost--free massive gravity Lagrangian~\eqref{mastergravitylag} and set the cosmological constant $\Lambda_c=0$ so as to have Minkowski as a vacuum solution.
When splitting the fields $\phi^\alpha=x^\alpha+A^\alpha+\eta^{\alpha\beta}\p_\beta \chi$ and the metric as $g\mn=\eta\mn+h\mn/\mpl$, at the linear level, the only place where the kinetic term for $\chi$ enters is through its coupling with $h\mn$. Symbolically, this is given by
\ba
\L_{\rm GFMG}^{(2)} \sim\,  h \p^2 h-m^2 \mpl^2 F\mn^2[A] +m^2 \mpl h^{\mu\nu}\( \Box \chi \eta\mn  \! - \! \p_\mu \p_\nu \chi \) +\frac{1}{\mpl}h\mn T^{\mu\nu}[\psi_i]\,,
\ea
where $T^{\mu\nu}[\psi_i]$ is the stress--energy tensor of the external fields $\psi_i$ coupled to gravity.
The mixing term can be taken care of by performing the field space rotation, symbolically,  \mbox{$h\mn = \tilde h\mn + \tilde\chi\eta\mn$} with $\Lambda_3^3=m^2 \mpl$ and $\tilde \chi$ the canonically normalized helicity--0 mode, $\tilde \chi=\Lambda_3^3 \chi$, so that
\ba
\L_{\rm GFMG}^{(2)} \sim\,  \tilde h \p^2 \tilde h-m^2 \mpl^2 F\mn^2[A]-(\p \tilde \chi)^2  +\frac{1}{\mpl}\tilde h\mn T^{\mu\nu}+\frac{1}{\mpl}\tilde \chi T  .
\ea
{\it At the linear level}, the coupling between $\chi$ and any non--conformal matter $\tilde \chi T$ is insensitive to the graviton mass $m$ and does not vanish in the massless limit. This is of course at the origin of the well--known {\it linear vDVZ--discontinuity} and its resolution lies in the {\it nonlinear interactions} which become increasingly important in the small mass limit as pointed out by A.~Vainshtein in~\cite{Vainshtein:1972sx}. In the context of nonlinear massive gravity the implementation of this  Vainshtein mechanism was considered for instance in~\cite{Deffayet:2001uk,Berezhiani:2013dw,Berezhiani:2013dca,Babichev:2013usa}. At the nonlinear level the theory involves interactions of the form $h (\p^2 \tilde \chi)^{n+1}/\Lambda_3^{3n}$, which implies that the theory is strongly coupled at the scale $\Lambda_3$~\cite{deRham:2010gu,deRham:2010ik}.\\

\subsection{Linearized theory on AdS}

\para{``$\Sigma$--model"}
When applied to AdS, the previous analysis has a rather different outcome:  Consider again the Lagrangian $\L_2$ in~\eqref{eq:L2} in its own right (\ie separated from its gravitational context) in $N=D$ dimensions but on an AdS spacetime,  so that the tensor $K$ and $X$ now read
\ba
\label{eq:KandXAdS}
K\mupn \equiv \delta\mupn - X\mupn,~~~ {\rm with}~~X\mupn = \sqrt{g^{\mu\ri}\pd_\ri \phi^\ai \pd_\nu \phi^\bi \gamma^{\rm (AdS)}_{\ai\bi}},
\ea
where $ \gamma^{\rm (AdS)}_{\ai\bi}$ is the AdS metric with curvature $ L^{-2}$, so that its associated Ricci tensor is \mbox{$R_{\ai\bi}=- L ^{-2}\gamma^{\rm (AdS)}_{\ai\bi}$.} Then the AdS curvature is sufficient to break the $U(1)$ gauge symmetry already at the linear level. Indeed at the linear level about the trivial vacuum  $g\mn= \gamma^{\rm (AdS)}\mn+h\mn/\mpl$, $ \phi^a=x^a+A^a$,
the AdS equivalent of~\eqref{eq:L2} reads
\ba
\L_2[\phi^a,\gamma^{\rm (AdS)}_{ab}] &=& \sqrt{-\gamma^{\rm (AdS)}}  \( -\frac 1{2}F\mn^2[A] -\frac{1}{2 L ^2}A_\mu^2  \) +\mc{O}(h_{\mu\nu}/\mpl)\,,
\ea
where all the contractions and covariant derivatives are with respect to the AdS spacetime metric. The appearance of a mass term for $A_\mu$ on AdS implies that the theory enjoys no accidental $U(1)$ and the helicity--0 mode $\chi$ acquires a kinetic term $A_\mu^2\supset (\p \chi)^2$.
It follows that on AdS the trivial vacuum $\phi^a=x^a$ is a perfectly well defined  and acceptable vacuum for the sigma model~\eqref{eq:L2} of $N=D$ fields, out of which $D-1$ are dynamical. Naturally, this result holds true for any generalization of that model $\L_2+\sum_{n=3}^D \alpha_n \L_n$.\\

\begin{figure}
  \centering
  \includegraphics[width=0.5\textwidth]{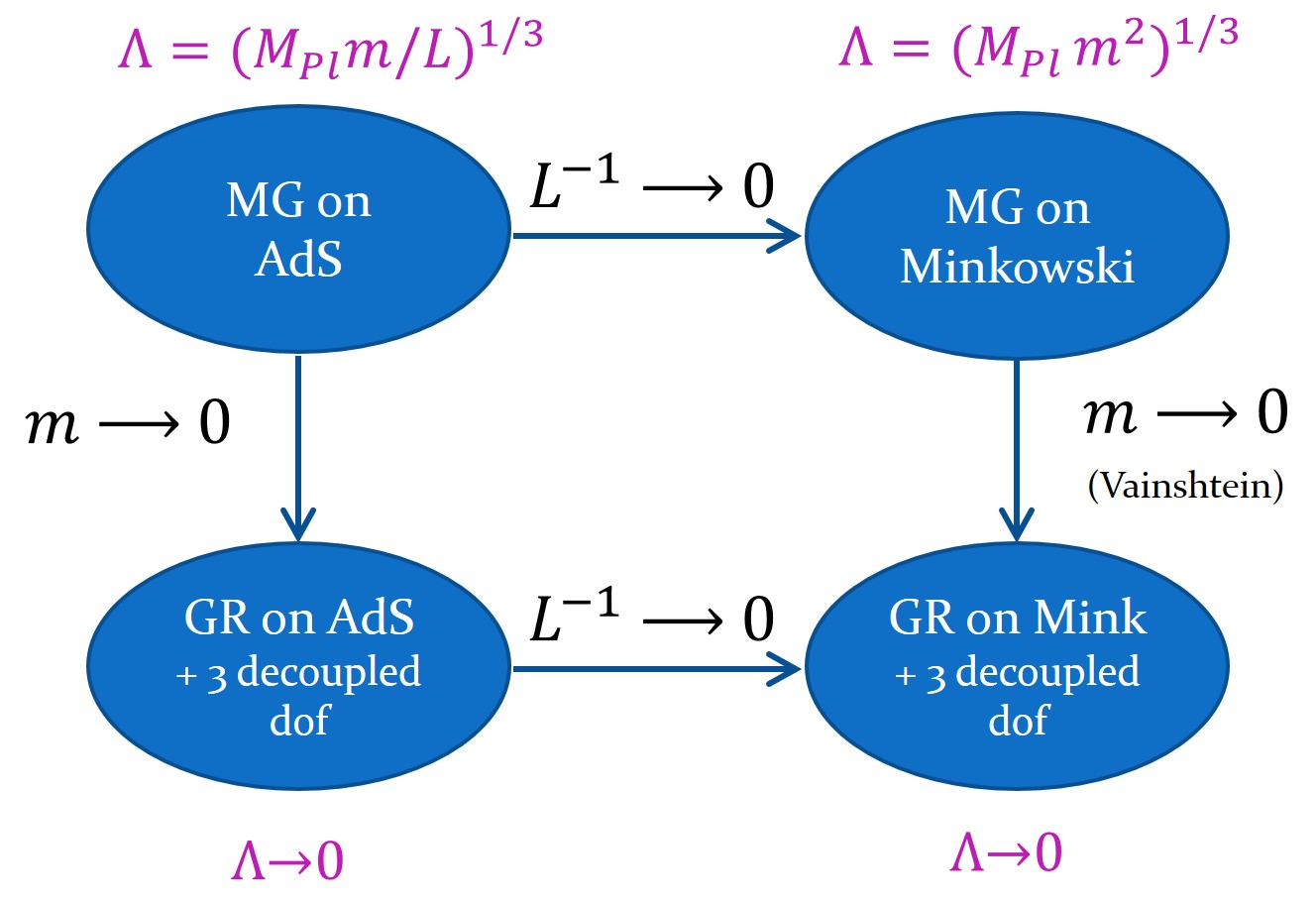}\\
  \caption{Massless limit of massive gravity (MG) on Minkowski and AdS. $L$ is the AdS scale and $\Lambda$ is the strong coupling scale.}\label{Fig:Limits}
\end{figure}

\para{Implications for massive gravity on AdS}
This result propagates to the case of gravity where it was  shown that the linearized vDVZ is absent on AdS ~\cite{Kogan:2000uy,Porrati:2000cp,Karch:2001jb,Porrati:2003sa,Porrati:2004mz}.
 Indeed, in the limit where the AdS curvature is larger than the graviton mass $m\ll L ^{-1}$, the canonically normalized field is now $\tilde \chi=\Lambda_*^3 \chi$ with
\ba
\label{eq:Lambda*}
\Lambda_*^3=\frac{\mpl m}{ L }\,,
\ea
and the coupling between $\tilde \chi$ and matter now goes as
\ba
\L_{ \tilde \chi T,\, {\rm AdS}}^{(2)}= \frac{m  L }{\mpl}\tilde \chi T \, \xrightarrow{m\to 0} \, 0\,,
\ea
which makes the massless limit of the linearized theory well--defined already at the linear level about AdS. This massless limit seems to occur without the need of a Vainshtein mechanism but we stress that
\begin{enumerate}
\item The Vainshtein mechanism is actually (secretly) active through the AdS background and this absence of discontinuity is in fact a direct implementation of the Vainshtein mechanism.
\item Strong coupling is still present in that theory. Indeed, the nonlinear theory includes interactions of the form $(\p \tilde \chi)^2 (\p^2 \tilde \chi)^{n-1}/\Lambda_*^{3n}$ implying that the theory is then strongly coupled at the scale $\Lambda_*$ as given in~\eqref{eq:Lambda*}.
\end{enumerate}
As shown in Fig.~\ref{Fig:Limits}, taking the limit $m\to 0$ and $ L^{-1}\to 0$ leads to the same scaling as if one had started straight from massive gravity on Minkowski and taken the massless limit. However, for a finite mass $m$ the strong coupling scale can be pushed higher if the AdS curvature is sufficiently large $m\ll L ^{-1}$, although  this comes at the price of working about a non--Minkowski reference metric.

In what follows we will show how one can capture some of these features of massive gravity on AdS (namely the absence of linearized vDVZ--discontinuity and a higher strong coupling scale) while maintaining  the reference metric nearly Minkowski. What we will consider instead is a non--trivial Lorentz--violating vacuum for the \stu fields.

\section{Nonlinear Hamiltonian analysis}
\label{sec:nha}

In the rest of this manuscript, we focus on the case where 
\be
N=D, ~~ ~~ f_{\mu\nu}=\eta_{\mu\nu}
\ee
and no longer distinguish between spacetime and target space indices. In this section we run the Dirac--Bergmann algorithm for the nonlinear theory~\eqref{MG-NLS}. We will see rigorously that even when decoupling gravity, the BD ghost is eliminated, as argued above, and in general there is no gauge symmetry to further reduce the number of DoFs in the fields $\phi^\alpha$.  Lorentz--invariant vacua are hence special as they re--introduce an accidental $U(1)$--symmetry at linear order, but that $U(1)$ is not a symmetry of the full sigma model and does not survive at higher order. Therefore in $D$ dimensions, $\phi^\alpha$ involves $D-1$ dynamical DoFs.\\

For simplicity, and without loss of generality, we focus in this section on the minimal model
\be
\L_1\sim - {\rm Tr}  X \sim - {\rm Tr} \sqrt{\eta^{-1}\pd \phi\eta \pd \phi^T}
\ee
given in~\eqref{MG-NLS}. The general model yields the same result.
To explicitly perform the Hamiltonian analysis, it is convenient to work with an equivalent form of the minimal Lagrangian:
\be
\label{lagl}
\mc{L}^{{\rm min}}_{\rm NLS} = - \frac12 \bar{\lambda}^{\mu\nu} \pd_\mu \phi^\ai \pd_\nu \phi^\bi \eta_{\ai\bi} - \frac12 \lambda_{\ai\bi}\eta^{\ai\bi}\,,
\ee
where the auxiliary variable $\lambda\mn$ is a symmetric tensor with inverse $\bar{\lambda}^{\mu\nu}$. See Appendix~\ref{sec:equivLag} for the equivalence between this Lagrangian and $ - {\rm Tr}X$. 
To derive the Hamiltonian, we perform an ADM--like split for the symmetric tensor $\lambda\mn$
\be
\lambda\mn =
\begin{pmatrix}
  -\lambda_0 + \mu^k \mu_k     & \ \ \mu_j  \\
  \mu_i & \ \  \sigma_{ij}
\end{pmatrix}   ,
\ee
where latin indices are for now lowered or raised with $\sigma_{ij}$ or its inverse $\sigma^{ij}$ respectively. The conjugate momenta for $\phi^\ai$ and $\sigma^{ij}$ are defined as
\bal
\pi_\ai &= \frac{\pd \mc{L}^{{\rm min}}_{\rm NLS}}{\pd \dot{\phi}^\ai} =  \frac1{\lambda_0}\dot{\phi}_\ai-\frac{\mu^i}{\lambda_0}\pd_i \phi_\ai ,
\\
\pi_{ij}&=  \frac{\pd \mc{L}^{{\rm min}}_{\rm NLS}}{\pd \dot{\sigma}^{ij}} = 0  ,
\label{eq:piij}
\eal
where the Lorentz index $\alpha$ is lowered with $\eta_{\ai\bi}$. After the Legendre transform, the Hamiltonian becomes quadratic in $\mu_k$ and linear in $\lambda_0$.  Integrating out $\mu_k$, we get
\bal
\mc{H}^{{\rm min}}_{\rm NLS} &= \mc{H}_0   +2 \lambda_0 \, \mc{C}^{(1)}  +  \mu^{ij} \mc{C}^{(1)}_{ij}\,,
\eal
where we have introduced the new set of Lagrange multipliers $\mu^{ij}$ to impose the relation~\eqref{eq:piij}, and we have defined
\bal
\mc{H}_0 &= \frac{1}2 \sigma^{ij} \pd_i\phi^\ai \pd_j \phi^\bi(\eta_{\ai\bi}+\pi_\ai\pi_\bi)
+\frac12 \sigma_{ij}\delta^{ij}   ,
\\
\mc{C}^{(1)} &= \pi_\ai \pi^\ai +1=0   ,
\\
\mc{C}^{(1)}_{ij} &=\pi_{ij}=0    ,
\eal
where $\mc{C}^{(1)}$ and $\mc{C}^{(1)}_{ij}$ are primary constraints. If now one further integrates out $\sigma^{ij}$, one can see that it is not possible to have any further constraints apart from the secondary associated with $C^{(1)}$.  But to be prudent, we show this explicitly by keeping $\sigma^{ij}$.\\

Since $\mc{C}^{(1)}$ and $\mc{C}^{(1)}_{ij}$ contain only conjugate momenta but not the fields themselves, it is clear that we have
\bal
\{ \mc{C}^{(1)}(\bfx), \mc{C}^{(1)}(\bfy) \} &= 0   ,
\\
\{ \mc{C}^{(1)}(\bfx), \mc{C}_{ij}^{(1)}(\bfy) \} &= 0   ,
\\
\{ \mc{C}_{ij}^{(1)}(\bfx), \mc{C}_{kl}^{(1)}(\bfy) \} &= 0   ,
\eal
and thus the time preservation of $\mc{C}^{(1)}$ and $\mc{C}^{(1)}_{ij}$ generate secondary constraints
\bal
\mc{C}^{(2)} &= \sigma^{ij}\pd_i \phi^\ai \pd_j \pi_\ai =0   ,
\\
\mc{C}_{ij}^{(2)} &= -\pd_i \phi^\ai \pd_j \phi^\bi (\eta_{\ai\bi}+\pi_\ai \pi_\bi) +\sigma_{ik}\delta^{kl}\sigma_{lj} =0   .
\eal
Then we check whether the time preservation of $\mc{C}^{(2)}$ and $\mc{C}^{(2)}_{ij}$ give rise to any tertiary constraints. Making use of the Poisson brackets:
\bal
\{ \mc{C}^{(2)}(\bfx) , \mc{C}^{(1)}(\bfy) \} &= 2\pd_i\pi_\ai\pd^i\pi^\ai(\bfx) \delta^{D-1}(\bfx-\bfy)  ,
\\
\{ \mc{C}^{(2)}(\bfx) , \mc{C}_{ij}^{(1)}(\bfy) \} &= \pd_{(i}\phi^\ai\pd_{j)}\pi_\ai(\bfx) \delta^{D-1}(\bfx-\bfy)   ,
\\
\{ \mc{C}_{ij}^{(2)}(\bfx) , \mc{C}^{(1)}(\bfy) \} &= -4\pd_{(i}\phi^\ai\pd_{j)}\pi_\ai(\bfx) \delta^{D-1}(\bfx-\bfy)   ,
\\
\{ \mc{C}_{ij}^{(2)}(\bfx) , \mc{C}_{mn}^{(1)}(\bfy) \} &= -2 \delta^{kl} \sigma_{l(j} \sigma_{i)(m} \sigma_{n)k} (\bfx)\delta^{D-1}(\bfx-\bfy)  ,
\eal
the consistency equations $\dot{\mc{C}}^{(2)}(\bfx)  = 0$ and $\dot{\mc{C}}_{ij}^{(2)}(\bfx)  = 0$ lead to
\bal
4\pd_i\pi_\ai\pd^i\pi^\ai~&\lambda_0~ &\hspace{-30pt}+ \pd_{(m}\phi^\ai\pd_{n)}\pi_\ai ~&\mu^{mn}
\hspace{-25pt}&=&  -\int \ud^{D-1} y \{ \mc{C}^{(2)}(\bfx) , \mc{H}_0(\bfy) \}   ,
\\
 -8\pd_{(i}\phi^\ai\pd_{j)}\pi_\ai ~&\lambda_0~ &\hspace{-30pt}-2 \sigma_{l(j} \sigma_{i)(m} \sigma_{n)k}\delta^{kl} ~&\mu^{mn}
\hspace{-25pt}&=& -\int \ud^{D-1} y \{ \mc{C}_{ij}^{(2)}(\bfx) , \mc{H}_0(\bfy) \}   ,
\eal
where $\{ \mc{C}^{(2)}(\bfx) , \mc{H}_0(\bfy) \} \neq 0$ and $\{ \mc{C}_{ij}^{(2)}(\bfx) , \mc{H}_0(y) \} \neq 0$ in general. This is a non--degenerate system of linear equations for unknowns $\lambda_0$ and $\mu^{mn}$.\\

One can indeed check that all $\lambda_0$ and $\mu^{mn}$ are determined by this system of linear equations. This is more easily performed in a  specific number of dimensions. For example, in $D=4$ dimensions,  one can show that the rank of the system of linear equations is 7, which corresponds to the number of $\lambda_0$ and $\mu^{mn}$. Thus, all $\lambda_0$ and $\mu^{mn}$ are determined. The Dirac--Bergmann algorithm ends here and all constraints are second class. Counting the phase space DoFs, we have, in $D=4$ dimensions,
\be
(4+6)\times 2 - (6+1) - (6+1) = 6=3\times2   ,
\ee
meaning that the number of physical DoFs is indeed 3.  This result was proven for the minimal model $\L_1$, but by continuity it holds for a general theory of~\eqref{MG-NLS}. We will re--confirm this result with a couple of different methods in the following.

\section{Exact non--trivial vacuum solution}
\label{sec:NonTrivialVacua}

Having shown that the massive gravity nonlinear sigma model also propagates two constraints that remove the BD ghost, and thus has 3 DoFs on generic backgrounds in $D=4$ dimensions, we shall now present an explicit example  where this occurs. In order to separate ourselves from the precise matter content of the model we work in the vacuum. In this sense our approach is different from, say, massive gravity on AdS, which requires a negative cosmological constant to source the background configuration. For the sake of simplicity, we focus once again on the minimal model, although our conclusions remain the same for any linear combinations of the Lagrangians $\L_n$.\\

\subsection{Plane--waves}

One of the difficulties in solving this equation for generic configurations of the fields $\phi^a$ lies in evaluating the square--root that enters in $X\mupn$. In what follows we will evaluate this square--root by performing perturbative expansions about the trivial vacuum, but for now we may consider the particularly simple --yet instructive-- example of  plane waves\footnote{Despite the terminology these solutions do not need to exhibit an oscillator behavior and the functions $F^I$ and $G^I$ are arbitrary.}. Take for instance
\ba
\label{eq:PlaneWave}
\bar \phi^\mu=x^\mu + \(F^I(t-x)+G^I(t+x)\)\delta^\mu_I \,,
\ea
where we have used the notation $x^0=t$, $x^1=x$ and the index $I$ labels the orthogonal directions, $I=2, \cdots, D-1$. This solves the vacuum equations of motion for arbitrary combinations of the Lagrangians $\L_n$ defined in~\eqref{LN} and  for arbitrary analytic functions $F^I$ and $G^I$. Indeed the tensor $\bar X\mupn$ associated with these plane wave configurations~\eqref{eq:PlaneWave} satisfies $\p_\mu  (\bar X^n)\mupn=0$ and $\p_\mu \({\rm Tr}\bar X^n\)=0$ no matter what the power $n$ is. This implies that the background configuration~\eqref{eq:PlaneWave} satisfies the equations of motion for the fields $\phi^\alpha$ for arbitrary combinations of the Lagrangians $\L_n$.\\

For instance, without loss of generality, we can set $G^I=0$ for any $I=2,\cdots,D-1$, $F^I=0$ for any $I=3,\cdots,D-1$ and write $F^{2}(t-x)=F(t-x)$. Then, if for simplicity, we work in $D=3$--dimensions and have
\be
\label{eq:simplevacuum}
\bar{\phi}^\mu=x^\mu + \( 0, 0, F(t-x) \)\,.
\ee
While the square root matrix $X\mupn$ has many branches of solution, it is understood that one should choose the branch that connects with the identity matrix when $F(t-x)\to 0$. So the matrix $\bar X\mupn$ associated with the non--trivial vacuum~\eqref{eq:simplevacuum} is
\be
\label{eq:bar X}
\bar{X}\mupn = \left(
\begin{array}{ccc}
 1-\frac{3}{8}  F'^2 & \frac{3}{8} F'^2 & -\frac{1}{2} F' \\
 -\frac{3}{8}  F'^2 & 1+ \frac{3}{8} F'^2 & -\frac{1}{2} F' \\
 \frac{1}{2} F' & -\frac{1}{2} F' & 1 \\
\end{array}
\right)\,,
\ee
where the prime denotes a derivative with respect to the function's argument, and one can indeed check that this matrix satisfies ${\rm Tr}[\bar X^n]=3$ for any power $n$, and so we have $\p_\mu \({\rm Tr}\bar X^n\)=0$. Furthermore, we can explicitly check that $\p_\mu \bar X\mupn=\p_\mu  \(\bar X^2\)\mupn=\p_\mu  \(\bar X^3\)\mupn=0$, so~\eqref{eq:bar X} satisfies the vacuum equations of motion for arbitrary combination of Lagrangians $\L_1+\alpha_2 \L_2+\alpha_3 \L_3$. This result is independent of the number of dimension and remains valid for arbitrary configurations of the form~\eqref{eq:PlaneWave}.

\subsection{Degrees of freedom}

Having established that the plane wave configurations~\eqref{eq:PlaneWave} are exact vacuum solutions,
we now proceed to evaluate the number of perturbative DoFs.
To establish the number of DoFs on that vacuum, it is sufficient to look at fluctuations of the form
\be
\phi^\ai = \bar{\phi}^\ai +\varepsilon V^\ai\,,
\ee
where we introduced a dimensionless parameter $\varepsilon$ to count the order in perturbations.
Focusing on the minimal model $\mc{L}_1$, then to quadratic order in $V$ (quadratic order in $\varepsilon$), we have
\be
\mc{L}_1 = \mc{F}(F')^{\mu\nu\ai\bi} \pd_{\mu}V_\nu \pd_{\ai}V_{\bi}\,,
\ee
where $\mc{F}^{\mu\nu\ri\si}$ are functions of $F'$.\\

The Hamiltonian analysis performed in \S~\ref{sec:nha} confirms that this model only has $D-1$ DoFs. About the trivial vacuum $\bar \phi^\ai=x^\ai$ ($F\equiv0$), $V^0$ is indeed an auxiliary variable. On more generic vacua, the auxiliary variable is instead a linear combination of the fields $V^\mu$, and to simplify the derivation we can perform a rotation in field space $V^\mu = W^\mu +R\mupn W^\nu$ so that $W^0$ is identified as the appropriate auxiliary variable. In $D=3$--dimensions, the appropriate rotation is given by
\ba
\label{eq:fieldRot}
V^0 = W^0, ~~~   V^i = W^i + R^i W^0\,,
\ea
with
\be
R^1= \frac{F'^2}{8+F'^2}\quad{\rm and}\quad {R^2}= \frac{4 F'}{8+F'^2}\,,
\ee
so that $\dot{W}^0$ entirely disappears from the resulting Lagrangian and there are only two conjugate momenta given by:
\be
\pi_i = \frac{\pd \mc{L}_{1}}{\pd \dot{W}^i}\,.
\ee
The Hamiltonian is then (to quadratic order in $\varepsilon$)
\bal
\mc{H}_{1} = \sum_i \pi_i \dot{W}^i  - \mc{L}_{1} = \mc{A}_2+ W^0 \mc{A}_1+ (W^0)^2 \mc{A}_0  ,
\eal
where $\mc{A}_{n}$ are functions of the background configuration $F'$ and are $n^{\rm th}$ order in the remaining phase space variables $W^{i}, \pi_i$. The exact expressions for $\mc{A}_0$ and $\mc{A}_1$ are given in~\eqref{eq:A2} and~\eqref{eq:A1} of appendix~\ref{sec:appendixPlaneWaves} but are irrelevant to this discussion.  $\mc{A}_0$ is given by
\ba
\mc{A}_0 =  \frac{128 F''^2}{\left(F'^2+8\right)^2 \left(3 F'^2+16\right)}\,,
\ea
and vanishes on the Lorentz--preserving vacuum where $F\equiv0$. About this trivial vacuum, $W^0$ is a Lagrange multiplier that generates a first--class constraint associated with an accidental $U(1)$--symmetry. Here we see explicitly that this symmetry is broken on generic backgrounds and while $W^0$ is still an auxiliary variable, it no longer generates a constraint for the phase space variables  $W^{i}, \pi_i$. Then all the $D-1$ remaining DoFs are dynamical and the resulting Hamiltonian (after integrating out the auxiliary variable $W^0$) is given by\footnote{For the trivial vacuum where $F\equiv 0$, one has $\mc A_1=\p_i \pi^i$ and $\mc A_0 \to 0$. This means that deviating from the surface $\mc A_1\ne 0$ would cost an infinite amount of energy and the fields are forced to live on the constrained surface where $\mc A_1=\p_i \pi^i\equiv 0$. However, as soon as $\mc A_0\ne 0$, one is allowed to deviate from that surface, and this deviation is encoded by the existence of an additional DoF.}
\be
\mc{H}_1 = \mc{A}_2 - \frac{\mc{A}^2_1}{4\mc{A}_0} \,.
\ee
This provides an explicit example of vacuum where all the expected DoFs are excited as they should.
Unfortunately, in this specific example, $\mc{A}_0>0$ and the resulting Hamiltonian is not bounded from below. As a result, in this specific example,  the background solution turns out to be  unstable. However, it represents an explicit proof--of--principle that non--trivial vacua can excite all the dynamical DoFs without needing to resort to a mixing with the tensor (gravitational) fields. In what follows we will show how to construct a more general class of stable vacua by considering solutions for the \stu fields  which are perturbative about the trivial one. We emphasize that looking for perturbative vacua is only used as an approximate tool to derive explicit vacua, but the theory also contains much more general classes of vacua.

\section{General perturbative backgrounds}
\label{sec:gpb}

We now present a different way to derive an acceptable non--trivial vacuum by relying on a perturbative approach.  This will allow us to derive the  Hamiltonian for a large class of vacua, confirming the DoF counting result of the full Hamiltonian analysis in \S~\ref{sec:nha} and \ref{sec:NonTrivialVacua}, and determining the absence of  ghosts and gradient instabilities  for a subclass of these vacua.

\subsection{Hamiltonian of fluctuations}
\label{sec:hamflu}

As considered previously, we look at fluctuations $V^\mu$ in a non--trivial vacuum $\bar \phi^\mu$,
\be
\label{phiexpBV}
\phi^\mu =\bar \phi^\mu +  \varepsilon V^\mu\,,
\ee
where as before $\varepsilon$ is a small dimensionless parameter which keeps track of the order in perturbations about the vacuum $\bar \phi^\mu$. Now for convenience and ease of the presentation,  the vacuum configuration itself is treated perturbatively,
\ba
\bar \phi^\mu=x^\mu+\ep \bar B^\mu\,,
\ea
and we will be considering the background to be perturbative in the dimensionless parameter $\ep$ (in what follows `barred' quantities will represent quantities that only involve the background).
For concreteness,  we focus on  a specific Lagrangian in what follows  and choose
\be
\label{K2lag}
\mc{L}^{\ai_2}_{\rm NLS} = 2  K^\mu_{\, [\mu} K^\nu_{\, \nu]} =  (K^\mu_{\, \mu} K^\nu_{\,  \nu} - K^\mu_{\, \nu} K^\nu_{\, \mu})\,,
\ee
with $K^\mu_{\, \nu}= \dd\mupn-X\mupn$, so $\mc{L}^{\ai_2}_{\rm NLS}$ differs from the minimal model $\L_1$ in~\eqref{MG-NLS}. Including higher $\alpha_n$ terms will add some computational complexity, but as we shall see below the $\ai_2$ term is sufficient for our purposes. In what follows we look at the Hamiltonian for the fluctuations $V^\ai$ living on top of the perturbed background $\bar\phi^\mu$. We therefore wish to compute the Hamiltonian quadratic in $\varepsilon$ and perturbatively in $\ep$. We will see that working  up to second order in $\ep$ is sufficient for this analysis.
The resulting quadratic Lagrangian for $V^\mu$ is given (symbolically) by
\be
\label{V2ndLag}
\mc{L}^{\ai_2}_{\rm NLS} = -\frac14  G_{\mu\nu}G^{\mu\nu}  + \ep\,  (\pd \bar B)^{\mu\nu\ri\si} \pd_{\mu} V_{\nu} \pd_{\ri} V_{\si} +\ep^2\,  (\pd \bar B \pd \bar B)^{\mu\nu\ri\si} \pd_{\mu} V_{\nu} \pd_{\ri} V_{\si} +\mc{O}(\ep^3)\,,
\ee
where we have defined
\be
G_{\mu\nu} \equiv 2 \pd_{[\mu} V_{\nu]}\,.
\ee
As expected, to lowest order in $\ep$, we recover the Maxwell term for $V^\mu$ and the theory enjoys an accidental $U(1)$--symmetry. The exact expressions at linear and quadratic order in $\ep$ in arbitrary dimensions are given in Appendix~\ref{sec:le}.\\

We now follow the same procedure as in the previous section, see Eq.~\eqref{eq:fieldRot}, and perform a field space rotation so as to identify the auxiliary variable $W^0$,
\be
\label{VtoW}
V^0 = W^0 ,~~~~ V^i = W^i + \bar T^i W^0\,,
\ee
and set the elements $\bar T^i$ perturbatively in $\ep$ so that the resulting Lagrangian does not involve any $\dot W^0$ (after appropriate integrations by parts). This procedure can be performed in arbitrary dimensions and if we focus for simplicity in $D=3$ dimensions, we get
\be
\label{Tiexp}
\bar T^i = \frac{\ep}{2} \bar F^{i0} + \frac{\ep^2}{8} \left( 2\dot{\bar B}^0\bar F^{0i} + \pd_j \bar B ^0\bar F^{ji} - 2\dot{\bar B}_j \pd^{(i}\bar B^{j)} -2\pd_j {\bar B}^0 \pd^{i}\bar B^{j}   \right) + \mc{O}(\ep^3)   ,
\ee
where we have defined
\be
\bar F^{\mu\nu} \equiv 2 \pd^{[\mu} \bar B^{\nu]}\,.
\ee
After substituting $\bar T^i$ into Eq.~(\ref{V2ndLag}), we can confirm that $W^0$ is manifestly an auxiliary variable.
To pass to the Hamiltonian formulation, we therefore define the conjugate momenta $\pi_i = {\pd \mc{L}}/{\pd \dot{W}^i}$
and get
\bal
\label{masHam}
\mc{H}^{\ai_2}_{\rm NLS} &= \frac12 \pi_i \pi^i + \frac14 G_{ij}G^{ij}  + \ep\, \mc{G}_1(\pi_i,W^i) + W^0 \left[ \pd_i \pi^i + \ep\, \mc{G}_2(\pi_i,W^i)  \right] - \ep^2 (W^0)^2  \bar{\mc{A}} + \mc{O}(\ep^3)  ,
\eal
where $\mc{G}_{1}$ and $\mc{G}_{2}$ do not depend on $W^0$ and their exact expressions is not relevant to the discussion here. In $D=3$ dimensions the term $\bar{\mc{A}}$ is given by
\be
\label{HW0}
\bar{\mc{A}} = - \frac18 \left( \dot{\bar F}_{ij}\dot{\bar F}^{ij} + 2 \pd_k \bar F_{ij} \pd^k \bar F^{ij} + 2 \dot{\bar F}^{0i}\pd^j \bar F_{ij}
- \pd_i{\bar F}^{0i}\pd_j {\bar F}^{0j} -  \pd_k \bar F_{0i} \pd^k \bar F^{0i}  \right)\,.
\ee
One important point to notice is that the term quadratic in the auxiliary variable $W^0$ only enters at quadratic order in $\ep$.
This means that up to leading and first order in the background expansion (zero and first order in $\ep$), the variable $W^0$ still acts as a Lagrange multiplier which generates the accidental $U(1)$--symmetry and removes one additional DoF. Indeed, had we truncated the theory to first order in $\ep$, $W^0$ would then act  as a Lagrange multiplier that enforces a primary constraint $\mc{C}^{(\ep)}_1=\p_i \pi^i+\ep \, \mc{G}_2\approx 0$ and  one can show that this constraint is first--class since it Poisson-commutes with itself
\be
\{\mc{C}^{(\ep)}_1(\bfx),\mc{C}^{(\ep)}_1(\bfy)\}=0   ,
\ee
the relevant part of $\mc{C}^{(\ep)}_1$ for this calculation being $\mc{C}^{(\ep)}_1\supset \pd_i \pi^i - \ep \dot{F}^{ij} G_{ij}/4$. Therefore, up to $\mc{O}(\ep)$, the  Hamiltonian (\ref{masHam}) still enjoys a gauge symmetry for any background $\bar B^\mu$ and only $D-2$ DoFs of $W^\mu$ are excited.\\

On the other hand, when the $\mc{O}(\ep^2)$ corrections are included, $\bar{\mc{A}}$ does not vanish for the background chosen and $W^0$ still remains an auxiliary variable but  ceases to be a Lagrange multiplier. To that order, integrating  out $W^0$ we then get
\be
\label{hamInt}
\mc{H}^{\ai_2}_{\rm NLS} =   \frac12 \pi_i \pi^i + \frac14 G_{ij}G^{ij} + \ep\, \mc{G}_1    + \frac{ \left( \pd_i \pi^i +\ep\,  \mc{G}_2   \right)^2}{4\ep^2 \bar{\mc{A}}}  \, .
\ee
Therefore, we can see that all the $D-1$ DoFs are now activated. The reason why the Hamiltonian is non--analytical in $\ep$ after integrating out $W^0$  is simply because our background itself is a perturbation around the trivial background $\phi^\ai=x^\ai$, where there is an accidental gauge symmetry and only $D-2$ DoFs are active. The non--analyticity in the Hamiltonian (\ref{hamInt}) reflects the fact that a DoF activated by a perturbative background is very weakly coupled, as we shall see more explicitly in what follows. It is straightforward to construct backgrounds for which $\bar{\mc{A}}$ does not vanish and is positive, and we shall construct approximate solutions below.\\

\subsection{The longitudinal mode}

In the last subsection, we have derived the quadratic Hamiltonian for the field $W^i$ on a generic background $\bar B^\mu$. Around the trivial background $\bar B^\mu=0$ (or $\bar \phi^\ai=x^\ai$), the longitudinal mode of $W^i$ is only a gauge mode. But, around a generic background (at least including the $\mc{O}(\ep^2)$ terms), this mode becomes dynamical and there are in total $D-1$ DoFs. Since the leading order ($\mc{O}(\ep^0)$) of the Hamiltonian (\ref{masHam}) is just the Maxwell theory, $D-2$ of these DoFs are just the transverse modes of an Abelian gauge field, thus totally free of ghost or gradient instabilities. Therefore, to study the linear stability of this theory, we only need to focus on the longitudinal mode $\pi_i \propto \pd_i \chi,~W_i \propto \pd_i \psi$.\\

From the Hamiltonian (\ref{hamInt}), we see that the leading contribution to the longitudinal momentum mode $\chi$ comes from the term ${ \( \pd_i \pi^i \)^2}/{4\ep^2 \bar{\mc{A}}}$. We shall scale it with $\ep$ so as to make the kinetic term of $\mc{O}(\ep^0)$:
\be
\label{normpi}
\pi_i=\ep \frac{\pd_i }{\nabla^2} \chi  ,
\ee
where $\nabla^2=\pd_i \pd^i$. Note that this is not yet the canonical normalization for the kinetic term, as there is still a characteristic scale in $\bar{\mc{A}}$.

Up to $\mc{O}(\ep)$ neither  $\mc{G}_1$ nor $\mc{G}_2$ contribute to the longitudinal mode $\psi$. This is because, up to $\mc{O}(\ep)$ in the Hamiltonian (\ref{masHam}), there is still a gauge symmetry, enforced by a first class constraint $\mc{C}^{(\ep)}_1$, as we mentioned above. To see this explicitly, note that, at order $\mc{O}(\ep)$, the contributions in $\mc{G}_1$ and $\mc{G}_2$ which are independent of $\pi^i$ are given by
\bal
\ep\, \mc{G}_1&\supset  -\frac{\ep}8 G^{ij} \left( \pd_k \bar B^k G_{ij} + 2 \bar F_{ij}\pd_k W^k + 2 \dot{\bar B}^0 G_{ij} \right) +\mc{O}(\ep^2) ,
\\
\ep \, \mc{G}_2 &\supset \ep \(  \frac14  \dot{\bar F}^{ij} G_{ij}  - \bar F^{0i}\pd^j G_{ij}\) +\mc{O}(\ep^2) \,.
\eal
These expressions are clearly independent of the longitudinal mode since $G_{ij}$ vanishes for the longitudinal mode $W_i\propto \pd_i \psi$. So the leading gradient terms, \ien, $\psi^2$ terms, come from the next order pieces in $\mc{G}_1$ and $\mc{G}_2$. Thus, to make the leading gradient terms of $\mc{O}(\ep^0)$, we can define the longitudinal mode as
\be
\label{normW}
W_i=\frac{1}{\ep}\frac{\pd_i }{\sqrt{\nabla^2}} \psi\,.
\ee
After performing the scaling of Eqs.~(\ref{normpi}) and (\ref{normW}), the leading contribution to the Hamiltonian for the longitudinal mode goes schematically as
\be
\mc{H}^{\rm L}_{\rm NLS} \sim  \frac{(\chi + \pd[(\pd \bar B)^2\pd \psi])^2}{(\pd^2 \bar B)^2} + (\pd \bar B)^2  (\pd \psi)^2 + \mc{O}(\ep)\,.
\ee
The first term always comes in as squared, so we may define
\be
\tilde{\chi} \sim \chi + \pd[(\pd \bar B)^2\pd \psi]  ,
\ee
and regard $\tilde{\chi}$ as the new conjugate momentum. Therefore, the leading Hamiltonian is
\bal
\label{hamleading}
\mc{H}^{\rm L}_{\rm NLS} &=   \frac{\tilde{\chi}^2}{4 \mc{\bar{A}}}  + \frac{1}{16} \bar F_{ij}\bar F^{ij} (\sqrt{\nabla^2} \psi)^2  + \frac12  \bar F_0{}^k \bar F_{0k} \frac{\pd_i\pd_j}{\sqrt{\nabla^2}}\psi \frac{\pd^i\pd^j}{\sqrt{\nabla^2}}\psi    + \frac14   \bar F_{0i} \bar F_{0j} \sqrt{\nabla^2}\psi \frac{\pd^i\pd^j}{\sqrt{\nabla^2}}\psi
\nn
 &  ~~~   -\frac34  \bar F_{0i} \bar F_{0j}  \frac{\pd^i\pd^k}{\sqrt{\nabla^2}}\psi  \frac{\pd^j\pd_k}{\sqrt{\nabla^2}}\psi    +  \mc{O}(\ep)  .
\eal
The linear stability of the longitudinal mode is guaranteed if one can find a background $\bar B^\mu$, such that $\mc{\bar A}$ is positive and the gradient term for $\psi$ is positive definite at least for a local patch of spacetime.

\subsection{Local backgrounds free of ghost and gradient instabilities}

For a smooth $\Lambda_2$--decoupling limit to be well--defined, it is essential that there are some stable background solutions in the massive gravity nonlinear sigma model. For the perturbative backgrounds being considered, we have come to the conclusion that the background is stable if the longitudinal mode is stable, that is, $\mc{H}^{\rm L}_{\rm NLS}$ is bounded from below. While one requires an exact solution to be stable across the whole spacetime, it is not necessary for a perturbative background to be stable globally, as the perturbative background may only be a good approximation of the underlying exact solution within a coordinate patch. Thus, to facilitate the stability analysis, we will expand a generic perturbative background within a local spacetime patch. Within this approach, it is easy to give explicit examples where ghost and gradient instabilities are both absent. \\

Suppose $\bar B^\ai$ has a characteristic length scale $L$, we can at least expect that within the spacetime patch $x<L$, $\bar B^\ai$ is smooth and analytical, and approximates the underlying exact solution to a sufficiently good extent. Thus, we Taylor expand $\bar B^\mu$ around the coordinate origin and substitute
\be
\bar B^\mu = \( \bar H^{\mu}_{\ri} x^\ri  +   \frac12 \bar M^{\mu}{}_{\ri\si} \frac{ x^\ri x^\si}{L} + \mc{O}\left( \frac{x^3}{L^2} \right)\)
\ee
into the Hamiltonian (\ref{hamleading}). Here $\bar H^{\mu}_{\ri}$ and $\bar M^{\mu}{}_{\ri\si}$ are constant. To leading order, both in $\ep$ and $x/L$, we have
\bal
\label{hamleadingx}
\mc{H}^{\rm L}_{\rm NLS} &=   \frac{\tilde{\chi}^2}{4 \bar{\mc{A}}}  + \frac{1}{16} \bar F_{ij}\bar F^{ij} (\sqrt{\nabla^2} \psi)^2  + \frac12  \bar F_0{}^k \bar F_{0k} \frac{\pd_i\pd_j}{\sqrt{\nabla^2}}\psi \frac{\pd^i\pd^j}{\sqrt{\nabla^2}}\psi    + \frac14   \bar F_{0i} \bar F_{0j} \sqrt{\nabla^2}\psi \frac{\pd^i\pd^j}{\sqrt{\nabla^2}}\psi
\nn
 &  ~~~   -\frac34  \bar F_{0i} \bar F_{0j}  \frac{\pd^i\pd^k}{\sqrt{\nabla^2}}\psi  \frac{\pd^j\pd_k}{\sqrt{\nabla^2}}\psi    +  \mc{O}\(\ep, \frac{x}{L} \)  ,
\eal
where now we have
\bal
\bar F_{\mu\nu}  &= 2 \bar H_{[\mu\nu]}  ,
\\
\label{AW0x}
\mc{\bar A} &= - \frac{1}{2L^2} \left(  \bar M_{[ij]0} \bar M^{[ij]}{}_{0} + 2 \bar M_{[ij]k} \bar M^{[ij]k} -2 \bar M_{[0i]0} \bar M^{[ij]}{}_{j} - \bar M_{[0i]}{}^{i} \bar M_{[0j]}{}^{j}- \bar M_{[0i]k} \bar M^{[0i]k} \right)  .
\eal
Now, since $\bar F_{\mu\nu}$ and $\mc{\bar A}$ are just constants, we can move $\pd_i$ and $\sqrt{\nabla^2}$ around by partial integration, so we may re--write Eq.~(\ref{hamleadingx}) as
\bal
\label{longHfin}
\mc{H}^{\rm L}_{\rm NLS} &=   \frac{\tilde{\chi}^2}{4 \mc{\bar A}}  + \frac{1}{16} \bar F_{ij}\bar F^{ij} \pd_k\psi \pd^k \psi  +   \bar F_{0i} \pd_j \psi  \bar F_0{}^{[i}  \pd^{j]} \psi    +  \mc{O}\(\ep, \frac{x}{L} \).
\eal
The gradient terms in this expansion are rather simple. In fact, they are manifestly positive definite. Thus, there are no gradient instabilities for any perturbative background within a local patch $L$. To determine the consistency of a perturbative background, one only needs to check for ghost instabilities, which amounts to checking whether or not a perturbative background gives rise to a positive $\mc{\bar A}$. \\

The equations of motion for $\phi^\mu$ in this approach becomes, to lowest and sufficient order,
\be
\bar M_{[\mu\nu]}{}^\mu = 0  ,
\ee
which should be satisfied by the perturbative background. Thus, $\mc{\bar A}$ can be viewed as a quadratic form of $\bar M_{\mu\nu\ri}$, subject to the constraint $\bar M_{[\mu\nu]}{}^\mu = 0$. Upon imposing this equation on Eq.~(\ref{AW0x}), one can show that there are positive directions in the Hessian of $\mc{\bar A}$ (viewed as a quadratic form of $\bar M_{\mu\ri\si}$). So there are perturbative background solutions that are free of ghost instabilities, and these backgrounds are the desired local $\Lambda_2$ backgrounds.  For a simple explicit example in $D=3$ dimensions, we note that one may choose $\bar M_{220}=1$ and $\bar M_{\mu\ri\si}=0$ for all others, then $\mc{\bar A} = 4>0$.\\

 In this section we have established the existence of stable vacua for the longitudinal mode. Since on this perturbative vacua, the other DoFs simply behave as an Abelian gauge theory (with small corrections), these DoFs are obviously free of ghost and gradient instabilities. Moreover, the longitudinal mode does not mix with the gauge modes to leading order. Thus, at least within our perturbative approach, there are backgrounds in the massive gravity nonlinear sigma model that are entirely free of ghost and gradient instabilities.

\section{The $\Lambda_2$--decoupling limit}
\label{sec:strong}

In \S~\ref{sec:l3limit}, we have seen that around the trivial background the longitudinal mode of ghost--free massive gravity only acquires a kinetic term via mixing with the tensor modes. Thus, around the trivial background, the theory is strongly coupled at the scale $\Lambda_3$. In the previous sections, we have shown that the massive gravity nonlinear sigma model~\eqref{MG-NLS} has $D-1$ DoFs and there are non--trivial backgrounds where all of these $D-1$ DoFs are excited and are stable, at least perturbatively. This means that on these generic vacua, ghost--free massive gravity ({\it i.e.}, the dRGT model) admits a  {\bf $\Lambda_2$--decoupling limit}:
\be
\label{l2lim}
\mpl \to \infty, ~~~~ m \to 0, ~~~~ \Lambda_2=\(\mpl ^{D}m^4\)^{\frac{1}{D+4}} \to {\rm fixed}  ,
\ee
which leads to
\be
S_{\rm GFMG} \to  \int\ud^D x \( - \frac{\mpl ^{D-2}}{4}{h}^{\mu\nu} \mc{E}^{\ri\si}_{\mu\nu} {h}_{\ri\si}  +  \Lambda_2^D   \sum_{n=2}^{D} \ai_n {K}^{\mu_1}_{[\mu_1}  {K}^{\mu_2}_{\mu_2} \cdots {K}^{\mu_n}_{\mu_n]} \)  ,
\ee
where $K\mupn=\dd\mupn- X\mupn$ with ${X}\mupn= \sqrt{\eta^{\mu\ri}\pd_\ri \phi^\ai \pd_\nu \phi^\bi \eta_{\ai\bi}}$.\\

We emphasize that directly setting $g_{\mu\nu}=\eta_{\mu\nu}$ in ghost--free massive gravity would be an inconsistent procedure. Rather, the correct way to obtain the massive gravity nonlinear sigma model is through the $\Lambda_2$--decoupling limit defined above. In this way, the healthy properties of ghost--free massive gravity can be carried over to the resulting scaled theory, \ien, the massive gravity nonlinear sigma model. To prove a smooth $\Lambda_2$--decoupling limit exists, we need to make sure the would--be decoupled theory has the right DoFs and there are backgrounds where these DoFs are well--behaved, which we have proven in the previous sections.  In what follows we can therefore work in this $\Lambda_2$--decoupling limit and determine how the strong couplings scale gets redressed by the scale $L^{-1}$.

\subsection{Generic operators}

In \S~\ref{sec:gpb}, we have shown that there are healthy backgrounds that are a small deviation from the trivial one $\bar \phi^\mu=x^\mu$. It may well be the case that there are healthy backgrounds far away from the trivial solution which could in principle be written as 
\be
g_{\mu\nu} = \eta_{\mu\nu} + h_{\mu\nu}/\mpl ,~~~~
\phi^\ai  = \bar{\phi}^\ai  + {V}^\ai = x^\ri {\bar Q}^\ai_\ri(x) + {V}^\ai   ,
\ee
where $\bar{\phi}^\ai$ is an {\it exact} background and $\bar Q_\ri^\ai\sim \mc{O}(1)$ is assumed to have a characteristic length scale $L$. One might also consider $\bar Q_\ri^\ai$ not to be $\mc{O}(1)$, but that simply amounts to redefining graviton mass $m$ and tuning dimensionless parameters $\ai_n$ (or $\bi_n$) away from $\mc{O}(1)$.\\

Schematically, the spacetime derivative of the background goes as
\be
\pd \bar{\phi} \sim  \pd (x\bar Q) \sim  \(1 + \frac{x}{L}\) \sim \mc{O}(1) ~~~{\rm within} ~~x\lesssim L  \,.
\ee
The matrix square root goes like $\mc{X} \sim \pd {\bar{\phi}} (1+\pd {V}/\pd {\bar{\phi}}+(\pd {V}/\pd {\bar{\phi}})^2 + \cdots)+ \mc{O}(h/\mpl)$.
Substituting these into the action (\ref{mastergravitylag}), the quadratic kinetic terms around this background are schematically given by
\be
S^{(\bar{\phi},k)}_{\rm GFMG}  = \int \ud^D x \left( - \frac{\mpl ^{D-2}}{4}{h}^{\mu\nu} \mc{E}^{\ri\si}_{\mu\nu} {h}_{\ri\si}  +  \Lambda_2^D  f{}^{\mu\nu}_{\ri\si} (\pd\bar{\phi})  \pd_\mu {V}^\ri  \pd_\nu {V}^\si    \right)\,,
\ee
where $f{}^{\mu\nu}_{\ri\si} (\pd\bar{\phi})$ represents functions of $\pd\bar{\phi}$.
Assuming all $\bi_n$ are $\mc{O}(1)$, we have
\be
f{}^{\mu\nu}_{\ri\si} (\pd\bar{\phi})  \sim \mc{O}(1)\,.
\ee
In our dimensional analysis below, we shall neglect all $\mc{O}(1)$ factors such as $f{}^{\mu\nu}_{\ri\si} (\pd\bar{\phi})$  as well as the Lorentz indices unless needed for the discussion. \\

As shown in the previous sections, one DoF in $V^\mu$ is not dynamical, so one can always perturbatively make a field redefinition
\be
V^\mu = F(\pd\bar\phi,W^\mu)\,,
\ee
so that $W^0$ is manifestly an auxiliary variable and the $D-1$ components of $W^i$ are dynamical. At linear order in $W^\mu$, this redefinition should reduce to a linear rotation similar to that of Eq.~(\ref{VtoW}) but with $\bar T^i$ now depending on the generic background $\bar\phi$. As shown in the previous section, the kinetic terms after the field redefinition will be schematically given by
\bal
\label{action2W}
S^{(\bar\phi,k)}_{\rm GFMG} \sim \int\ud^D x \(  - \mpl ^{D-2} h\mc{E} h + \Lambda_2^D \bigg[ (\pd W^i)^2 + W^0  \(  \pd \pd W^i  + L^{-1} \pd W^i \) + L^{-2}(W^0)^2 \bigg]\)\,.
\eal
There is a characteristic scale $L^{-1}$ coming out of the background every time a derivative is shifted from $W^\mu$ to the background $\bar\phi$. As $W^0$ is an auxiliary field, one can integrate it out, which, to leading order in perturbations in $W^\mu$, should be
\be
\label{W0leading}
W^0|_{\rm leading} \sim L^2 \pd^2 W^i + L\pd W^i\,.
\ee
We will later include all possible nonlinear terms of $W^\mu$ for $W^0$. Therefore, integrating out $W^0$ at leading order, we have
\bal
\label{action2Wint}
S^{(\bar\phi,k)}_{\rm GFMG} \sim \int\ud^D x  \( - \mpl ^{D-2} h\mc{E} h + \Lambda_2^D \bigg[ (\pd W_{\perp}^i)^2 +  L^2 \(  \pd^2 W_{\parallel}^i \)^2  \bigg]\)\,,
\eal
where $W_\perp^i$ represent the $D-2$ transverse modes and $W_\parallel^i$ the longitudinal mode which is absent on the trivial vacuum but not on generic ones. Note that in deriving Eq.~(\ref{action2Wint}) we have neglected the $L\pd W^i$ term of Eq.~(\ref{W0leading}). This is because a derivative on $W^\mu$ is greater than $L^{-1}$ within $x\lesssim L$, so one can symbolically think of $L\pd$ as a large number.\\

The canonical normalizations are then
\be
\hat{h} \sim \mpl ^{\frac{D-2}{2}} h ,~~~ \hat{W}_{\perp}^i \sim \Lambda_2^{\frac{D}{2}} W_{\perp}^i, ~~~ \hat{W}_{\parallel}^i \sim  \Lambda_2^{\frac{D}{2}} L\pd W_{\parallel}^i\,,
\ee
and from these normalizations, it is obvious that the lowest strong coupling scale should come from some pure $W^\mu$ interactions, \ien, terms without $h$.\\

Although the model is fixed (up to a few parameters), we now have the freedom to choose the vacuum $\bar\phi$. This choice will then affect the normalization and hence the scale of the interactions. We shall first assume that all  {\it a priori} conceivable terms exist, and then comment on specific classes of vacua  where certain terms happen to cancel. Before canonical normalization and integrating out $W^0$, a generic  interaction for $W^\mu$ is given by
\be
\label{genterm}
 \Lambda_2^D L^{-P} \pd^{T-P} (W^0)^Q (W^i)^{T-Q}, ~~{\rm with}~~T\geq 3, ~T\geq P\geq 0,~ T\geq Q\geq 0\,.
\ee
Next, we integrate out $W^0$, which, including all possible nonlinear orders, may be written as
\be
W^0 \sim \sum_{K,N}  L^{K}\pd^N (W^i)^{N-K} W^i,~~~{\rm with}~~N\geq K;~~ {\rm when}~N=K,~~K=1,2 ,
\ee
where we have used Eq.~(\ref{W0leading}) for $W^0$. (In here $N$ is not to be confused  with the dimension of the target space that appeared earlier.) Substituting $W^0$ into Eq.~(\ref{genterm}), a generic interaction term is then given by
\be
 \Lambda_2^D L^{-P+QK} \pd^{T-P+QN}  (W^i)^{T+Q(N-K)}\,.
\ee
Assuming that $M$ of the $W^i$ are the longitudinal mode $W^i_{\parallel}$ and the rest are the transverse mode $W^i_{\perp}$, the canonical normalization gives
\be
\frac{L^{QK-P-M} } {\Lambda_2^{\frac{D}2 (T-2+Q(N-K))}} ~ \pd^{T-P+QN-M} (\hat{W}_{\parallel}^i)^M (\hat{W}_{\perp}^i)^{T+Q(N-K)-M}\,,
\ee
with integers $T,N,K,P,Q,M$ satisfying
\be
\label{range1}
T\geq 3,~N\geq K, ~0\leq P\leq T,~ 0\leq Q\leq T,~ 0 \leq M\leq T+Q(N-K)\,.
\ee
For operators with $QK-P-M\leq 0$, the corresponding operator is either relevant or has a strong coupling scale that is no smaller than $\Lambda_2$ (simply noting that $T-2+Q(N-K)>0$).

\subsection{Strong coupling scale}

The operators that enter at the lowest energy scale satisfy $QK-P-M>0$, which requires
\be
K>0\,.
\ee
For these operators, the associated energy scale is a geometric mean of $\Lambda_2$ and $L^{-1}$ (the characteristic scale of the background):
\be
\Lambda_{2*}= (\Lambda_2^m L^{-n})^{1/(m+n)}
\ee
with $m=QK-P-M>0$ and $n=T+Q(N-K)-2>0$. For the stable perturbative backgrounds we have identified with the local coordinate expansion, the existence of a valid effective field theory requires that $L$ is larger than $\Lambda_{2*}^{-1}$, which implies $L^{-1}<\Lambda_2$.
It follows that the lowest interaction scale then comes from a geometric mean where $L^{-1}$ has as many powers as possible. That is, the lowest strong coupling scale corresponds to the greatest ratio of
\be
\label{rar}
\frac{m}{n} = \frac{QK-P-M}{T+Q(N-K)-2}   .
\ee
In summary, using the relation (\ref{range1}) as well as  $K>0$, it is clear that the greatest ratio corresponds to $N=K=2,~P=M=0,~Q=T$ with $T=3$. This ratio comes from cubic terms that go like
\be
 \Lambda_2^D \pd^3 (W^0)^3,~~{\rm with} ~~W^0 \sim L^2\pd^2W_{\perp}^i   .
\ee
Since $W^0$ is an auxiliary field, the $\pd^3$ in front of  $(W^0)^3$ should only contain spatial derivatives. Thus, if all {\it a priori} possible terms exist in the perturbative expansion of $W^\mu$ on some background $\bar\phi$, then the lowest strong coupling is given by
\be
\Lambda_{2*}^{\rm min} = (\Lambda_2^{{D}} L^{-12})^{\frac{1}{D+12}}\,  .
\ee
On the other hand, it is conceivable that for certain backgrounds some operators may not exist or cancel out. In addition, some operators may be removable by field redefinitions. Around those backgrounds, $\Lambda_{2*}$ can potentially be raised to
\be
\Lambda_{2*}^{\rm max} = (\Lambda_2^{D} L^{-4})^{\frac{1}{D+4}}\,,
\ee
which corresponds to the strong coupling scale for $T\to \infty$, \ien, vertices with a large number of legs. For example, for  backgrounds for which all $P\leq 3$ (or $M\leq 3$, or $P\leq 2$ and $M\leq 1$, etc) terms cancel out, we only have interactions with
\be
P+M\geq 4 \, .
\ee
Then the greatest ratio of $m/n$ is given by $2 -(P+M-4)/(T-2)$, which tends to $2$ when $T \to \infty$ and $P,M$ remain finite. \\

In summary, the precise value of the  strong coupling scale depends on the detailed properties of the vacuum and its characteristic scale $L$, which should be analyzed on a case by case basis. But the range of the dressed scale $\Lambda_{2*}$ is
\be
(\Lambda_2^{{D}} L^{-12})^{\frac{1}{D+12}}< \Lambda_{2*} < (\Lambda_2^{D} L^{-4})^{\frac{1}{D+4}}\,,
\ee
and can be parametrically larger than the standard $\Lambda_3$ scale one typically derives in massive gravity. Notice that when $L$ is so large that the resulting scale $\Lambda_{2*}$ becomes comparable or smaller than $\Lambda_3$ then the interactions with the gravity can no longer be ignored and the correct strong coupling scale does not actually fall below $\Lambda_3$. 

\section{$U(1)$ symmetry in  2D}
\label{sec:2d}

The general results of the previous sections apply to dimensions greater than two. In $D=2$ dimensions, the massive gravity nonlinear sigma model has an extra gauge DoF, on top of the constraints that eliminate the BD ghost. So there is no physical DoF in the 2D massive gravity nonlinear sigma model, if the internal space is of the same dimension as the spacetime.  In this section, we show explicitly  the gauge transformation around an arbitrary background.\\

The general massive gravity nonlinear sigma model in 2D is given by
\be
\mc{L}_{2D} = {\rm tr}\sqrt{\eta^{\mu\ri}\pd_\ri\phi^a\pd_\nu\phi^b\eta_{ab}}\,.
\ee
For simplicity, we adopt here a Euclidean signature for $\eta^{\mu\nu}$ and $\eta_{ab}$, as our goal is mainly to count the number of DoFs in the theory. Assuming $\bar A^\mu$ is a background solution which satisfies the equations of motion, we look for a small perturbation around it
\be
\phi^\mu = \bar A^\mu + \varepsilon V^\mu  .
\ee
The equations of motion for $\bar A^\mu$ are
\be
\label{Aeom4}
\pd_\nu \(\bar f \pd_\mu \bar A^\mu \) + 2  \pd^{\mu} \( \bar f \pd_{[\mu}\bar A_{\nu]} \)=0  ,
 ~~~{\rm with}~~~
  \bar f=\frac{1}{\sqrt{(\pd_\mu \bar A^\mu)^2 + 2 \pd_{[\mu}\bar A_{\nu]}\pd^{[\mu}\bar A^{\nu]}}} .
\ee
The quadratic Lagrangian for the perturbations  $V^\mu$ on the vacuum $\bar A^\mu$ is captured by
\be
\mc{L}_{2D}^{(2)} = \frac{\bar f^3}{2} \left( \pd_{[\mu} \bar A_{\nu]} \pd_\ai V^\ai -  \pd_\ai \bar A^\ri \pd_{[\mu} V_{\nu]}  \right)^2  \,,
\ee
where $\bar A_{\nu}$ satisfies the background equations of motion (\ref{Aeom4}). By direct calculation, one can show that this Lagrangian is invariant under the infinitesimal gauge transformation:
\be
V_\mu \to V_\mu + \pd_\mu \xi  -  \frac{ 2 \pd_{[\mu} \bar A_{\nu]}\pd^\nu \xi  }{ \pd_\ai \bar A^\ai  }  ,
\ee
where $\xi(t,x)$ is the gauge parameter, once the on--shell conditions are imposed on $\bar A^\mu$. This implies that the $U(1)$--symmetry remains about any on--shell background of the theory. Since we have worked at quadratic order about an arbitrary background, our analysis is equivalent to working to all orders about the trivial background. 
The helicity--0 mode is hence fully absent from the theory which propagates no physical degrees of freedom in $D=2$ dimensions. The existence of this symmetry is very specific to $D=2$ dimensions and as we have seen does not generalize to higher dimensions where the $U(1)$--symmetry is broken in the full theory.

\section{Discussions}
\label{sec:sum}

In this paper, we have developed the $\Lambda_2$--decoupling limit of Lorentz--invariant massive gravity (specifically ghost--free massive gravity~\cite{deRham:2010ik,deRham:2010kj}). This is an approximate description of a large family of solutions of Lorentz--invariant massive gravity, all of which spontaneously break Lorentz invariance. Hence this excludes the usual Lorentz invariant vacuum which lies within the $\Lambda_3$ regime. Interestingly the $\Lambda_2\gg \Lambda_3$ regime is far closer in spirit to the decoupling limit of massive gravity on AdS where the strong coupling scale is also parametrically higher.
As in the case of massive gravity on AdS, the vDVZ--discontinuity is simply absent already at the linear level, and hence these backgrounds  easily comply with existing tests of gravity. \\

Beyond the scheme of massive gravity, we have also shown an interesting connection between ghost--free massive gravity as a generalization of the $p$--brane Nambu--Goto action. In particular, we have pointed out that the ghost--free graviton potential can be viewed as a non--standard nonlinear sigma model that uniquely evades the compact requirement for the target space. This evasion is different from all the known examples where some auxiliary gauge trick is utilized and the first class constraints associated with the gauge symmetries explicitly project out the would--be ghost, while the massive gravity nonlinear sigma model makes use of second class constraints to project out the would--be ghost.\\

The uniqueness of ghost--free massive gravity, which essentially is due to the uniqueness of the matrix square root and anti--symmetrization scheme of the graviton potential, suggests that Lagrangian (\ref{MG-NLS}) is a unique generalization of the Nambu--Goto action that eliminates the ghost associated with the negative direction of the target space~\cite{deRham:2015ijs}. Without spoiling the spirit of this uniqueness, a further generalization is to promote the $\ai_n$ parameters to be functions of $\phi^A$, which also gives rise to a consistent nonlinear sigma model~\cite{deRham:2015ijs}. On the other hand, letting the target space have more than one negative direction, such as $(--,+\cdots+)$, is necessary problematic~\cite{deRham:2015ijs}. Such a nonlinear sigma model has more than one ghost in the spectrum, but the unique matrix square root and anti--symmetrization scheme can only eliminate one ghost\,\footnote{By definition, a nonlinear sigma model has only one target space. If one allows for more than one target space, there may be field theories that are ghost--free but with more than one negative internal direction.}. (In the Nambu--Goto special case, having more than one negative direction is possible as there are more than one diffeomorphism invariance, if $D=p+1>1$.)\\

For most of this manuscript,  we have restricted ourselves to an internal space which is at least as large as the spacetime dimension,  $N\ge D$. The case $N<D$ has its own interest, and was for example applied for the description of realistic condensed matter systems using the AdS/CFT correspondence in~\cite{Alberte:2014bua}. However, the absence of the BD ghost for $N<D$ is more subtle. As shown in~\cite{Alberte:2013sma}, in some cases of $N<D$, all the $N$ DoFs may propagate. We note that this happens whenever the lapse function squared of the reference metric $-f_{00}+f_{0k}(f^{-1})^{kl}f_{l0}$ vanishes, (here we have extended the target space metric $f_{AB}$ with zeros such that it formally has the same dimension as $g_{\mu\nu}$), which is when the unitary gauge Hamiltonian proof of the ghost--free--ness of massive gravity with a general reference metric~\cite{Hassan:2011tf} fails. \\

We have studied the massive gravity nonlinear sigma model by performing a nonlinear Hamiltonian analysis/Dirac--Bergmann algorithm, finding an exact solution and examining perturbations on that solution, and examining perturbations on a general perturbative background and determining its stability. Our study of the massive gravity nonlinear sigma model indicates that:
\begin{itemize}
\item
Ghost--free massive gravity ({\it i.e.}, the dRGT model) admits a smooth $\Lambda_2$--decoupling limit where the tensor modes are completely decoupled, and the whole matrix square root and anti--symmetrization structure is kept intact.
\item
Ghost--free massive gravity admits many non--trivial $\Lambda_2$--backgrounds that are stable, around which all the $D-1$ DoFs are propagating. These backgrounds need non--vanishing support from the vector modes, and spontaneously break the Lorentz invariance with the strength of the graviton Compton length scale.
\item
There is no linear vDVZ--discontinuity around these $\Lambda_2$ backgrounds. Thus these backgrounds trivially pass the local gravity tests such as the solar system tests for a Hubble scale graviton Compton length. In some sense, the $\Lambda_2$ backgrounds are the ones with the Vainshtein mechanism already implemented.
\item
Around these $\Lambda_2$ backgrounds, the strong coupling scale is raised to $\Lambda_{2*}$, which is parametrically larger than $\Lambda_3$.
\end{itemize}

It has been shown that homogeneous and isotropic cosmological solutions, as well as static, spherically symmetric black holes, in ghost--free massive gravity are absent/unstable~\cite{Berezhiani:2013dw, Berezhiani:2013dca, DeFelice:2012mx}, and it has been argued that the ``natural'' cosmological solutions in ghost--free massive gravity are inhomogeneous/anisotropic and the ``natural'' black hole solutions are non--static/spherically symmetric, the deviations from the exact symmetries being typically of $\mc{O}(m^2)$. In the $\Lambda_2$ decoupling limit, we are forced to break Lorentz symmetries in order to have stable backgrounds, and indeed we expect that it is the $\Lambda_2$ decoupling limit that is the most appropriate description of the generic inhomogenous cosmologies in massive gravity. We remind the reader that this forced inhomgeneity is not in conflict with observations since the scale of the inhomgeneity is set by $m^{-1}$ which can be made arbitrarily large, and is usually taken to be at least of the order of the current Hubble horizon. \\

The existence of the $\Lambda_2$--decoupling corresponds to a description of backgrounds which in unitary gauge will locally take the form
\be
\bar g_{\mu\nu} = \pd_\mu\phi^\ai\pd_\nu\phi^\bi\eta_{\ai\bi} + \mc{O}(m^2)  .
\ee
They are physically different solutions from the Minkowski metric $\eta_{\mu\nu}$ even if the $\mc{O}(m^2)$ corrections were excluded, and the differences will show up in perturbations in the gravitational sector. If $m^{-1}$ is taken to be a cosmological scale (of the order of the observable Universe today), all these backgrounds have essentially an approximately FRW geometry below the Hubble horizon, and at scales larger than the current Hubble scale can become inhomogeneous. We thus expect that the $\Lambda_2$ solutions describe a typical inhomogeneous cosmology, which may be approximately homogenous out to the scale $m^{-1}$. Once again, these $\Lambda_2$  backgrounds have the virtue that there is no linear vDVZ--discontinuity, and hence it will be significantly easier to satisfy current tests of gravity, raising the possibility that it is these $\Lambda_2$ backgrounds that may have the most direct connection with phenomenology.\\

We have shown that a $\Lambda_2$ background that is perturbatively away from the trivial $\Lambda_3$ background $\bar \phi^\ai=x^\ai$ is sufficient to excite the longitudinal mode. This suggests that one can continuously connect the trivial $\Lambda_3$ background with some nontrivial $\Lambda_2$ backgrounds. There may be some backgrounds such that in some local region (for instance around a star or black hole) the background is of the $\Lambda_2$ type, and asymptotically the background approaches the $\Lambda_3$ limit. How a particular background is chosen is determined by the initial and boundary conditions.\\

\vskip 10pt
\noindent {\bf Acknowledgments:}
We would like to thank Paul Saffin for helpful discussions.
CdR is supported by a Department of Energy grant DE-SC0009946.
AJT and SYZ are supported by Department of Energy Early Career Award DE-SC0010600.

\appendix

\section{Appendices}

\subsection{Equivalent Lagrangians for the minimal model}
\label{sec:equivLag}

In the Vielbein formulation of ghost--free massive gravity~\cite{Hinterbichler:2012cn}, the flat space limit of the minimal model (in the $X^\mu_\nu$ formulation) is given by
\bal
S_{\rm m} &\propto \int \epsilon_{\mu\nu\ri\si} \Lambda^\mu{}_\li \ud \phi^\li \wedge \ud x^\nu  \wedge \ud x^\ri  \wedge \ud x^\si
\propto \int \ud^4 x \Lambda^\mu{}_\nu \pd_\mu \phi^\nu   ,
\eal
where $\Lambda^\mu{}_\nu$ is an auxiliary field, satisfying $\Lambda^\mu{}_\nu \eta_{\mu\si} \Lambda^\si{}_\ri  = \eta_{\nu\ri}$. Thus, the minimal model Lagrangian can be written in either of the following equivalent forms
\bal
\mc{L}_{\rm m1} &= \Lambda^\mu{}_\nu \pd_\mu \phi^\nu + \lambda_{\ri\si} ( \Lambda^\ri{}_\ai \eta^{\ai\bi} \Lambda^\si{}_\bi -\eta^{\ri\si} )  ,
\\
\label{lag2Lmd}
\mc{L}_{\rm m2} &= \Lambda^\mu{}_\nu \pd_\mu \phi^\nu + \lambda^{\ri\si} ( \Lambda^\ai{}_\ri \eta_{\ai\bi} \Lambda^\bi{}_\si -\eta_{\ri\si} )  ,
\eal
where $\lambda_{\ri\si}$ and $\lambda^{\ri\si}$ are symmetric in exchanging $\ri$ and $\si$. Since $\Lambda^\mu{}_\nu$ is quadratic in either of the two Lagrangians, we can easily integrate it out respectively. Up to a global rescaling of $\lambda_{\ai\bi}$, we get
\bal
\label{lag1lambda}
\mc{L}_{\rm m1} &= - \frac12 \bar{\lambda}^{\ri\si} \pd_\ri \phi^\ai  \pd_\si \phi^\bi \eta_{\ai\bi}  -\frac12\lambda_{\ai\bi}\eta^{\ai\bi} ,
\\
\label{lag2lambda}
\mc{L}_{\rm m2} &= - \frac12 \eta^{\ri\si} \pd_\ri \phi^\ai  \pd_\si \phi^\bi \bar{\lambda}_{\ai\bi}  -\frac12 \lambda^{\ai\bi}\eta_{\ai\bi} ,
\eal
where $\bar{\lambda}^{\ri\si}$ is the inverse of $\lambda_{\ri\si}$. In \S~\ref{sec:nha}, we take advantage of Lagrangian (\ref{lag1lambda}), as this form entitles an ADM--like splitting for $\lambda_{\ai\bi}$ in the full Hamiltonian analysis. This action also resembles the Polyakov action to some extent. Expressions similar to Lagrangian (\ref{lag2lambda}), with gravitons activated, have been utilized to re--confirm the absence of the BD ghost in ghost--free massive gravity~\cite{Golovnev:2011aa,Hassan:2012qv}. Further integrating out $\lambda_{\ri\si}$, we arrive at
\bal
\mc{L}_{\rm m1} &= -{\rm tr} \sqrt{\eta^{-1}\pd \phi\eta \pd \phi^T} = -{\rm tr} \sqrt{\eta^{\mu\ri} \pd_\ri \phi^\ai\pd_\nu \phi^\bi \eta_{\ai\bi} },
\\
\mc{L}_{\rm m2} &= -{\rm tr} \sqrt{\pd \phi^T\eta^{-1} \pd \phi \eta} = -{\rm tr} \sqrt{\pd_\ri\phi^\mu \eta^{\ri\ai}\pd_\ai\phi^\bi \eta_{\bi\nu}}.
\eal

\subsection{Plane--wave Hamiltonian}
\label{sec:appendixPlaneWaves}
To count the DoFs about the non--trivial plane--wave vacuum configuration \eqref{eq:PlaneWave}, we work in the Hamiltonian formalism. To provide an explicit derivation, we focus on the $D=3$ dimensional case provided in Eq.~\eqref{eq:simplevacuum} and without loss of generality, we consider solely the Lagrangian $\L_1$. \\

We consider linear fluctuations $V^\ai$ about the vacuum configuration $\bar{\phi}^\ai$ so that the fields $\phi^\ai$ take  the form
\be
\phi^\ai = \bar{\phi}^\ai + V^\ai\,.
\ee
To quadratic order in fluctuations, we then have
\be
\mc{L}^{(2)}_1 = \mc{F}(F')^{\mu\nu\ri\si} \pd_{\mu}V_\nu \pd_{\ri}V_{\si}   ,
\ee
where $\mc{F}^{\mu\nu\ri\si}$ are functions of $F'$. Since the BD ghost is absent from this theory (as confirmed by the Hamiltonian analysis of \S~\ref{sec:nha}, some combination of the $V^\mu$'s must play the role of a Lagrange multiplier. On arbitrary backgrounds the Lagrange multiplier is a linear combination of the $V^\mu$'s, and, to make the primary constraint manifest, we can rotate the fluctuations $V^\ai$ in field space
\be
V^0 = W^0, ~~~   V^i = W^i + R^i W^0\,,
\ee
in such a way that $W^0$ becomes an auxiliary field. By requiring $\pd\mc{L}^{(2)}_{1}/\pd \dot{W}^0$ not to contain $\dot{W}^\mu$, we get in $D=3$ dimensions
\be
R^1= \frac{F'^2}{8+F'^2}\quad{\rm and}\quad {R^2}= \frac{4 F'}{8+F'^2}\,,
\ee
which gets rid of $\dot{W}^0$ completely (after appropriate integrations by parts while maintaining at most one time--derivative per field). Then one can define the conjugate momentum for $i=1,2$
\be
\pi_i = \frac{\pd \mc{L}^{(2)}_{1}}{\pd \dot{W}^i}
\ee
to pass to the Hamiltonian
\bal
\mc{H}^{(2)}_{1} = \sum_i \pi_i \dot{W}^i  - \mc{L}^{(2)}_{1} = \mc{A}_2+ W^0 \mc{A}_1+ (W^0)^2 \mc{A}_0   ,
\eal
where $\mc{A}_{n}$ are functions of the background configuration $F'$ and are $n^{\rm th}$ order in the remaining phase space variables $W^{i}, \pi_i$,
\ba
\label{eq:A0}
\mc{A}_0 &=&  \frac{128 F''^2}{\left(F'^2+8\right)^2 \left(3 F'^2+16\right)}  , \\
\label{eq:A1}
\mc{A}_1&=&
\frac{-4}{(F'^2+8)^3 (3 F'^2+16)}   \bigg[(F'^2+8)^2 F'' (\pd_2W^1 (F'^2-8)+4 (\pd_2W^2+\pd_1W^1) F'+8 \pd_1W^2)  \notag \\
&&
+2  (F'^2+8)(3 F'^2+16) \big( F' (\pd_2\pi_2 F'-4 \pd_2\pi_1+4 \pd_1\pi_2)
-8 (\pd_1\pi_1+\pd_2\pi_2)\big) \notag\\
&&
 +F'^4 (6 \pd_2\pi _2 +\pd_2 W^1 F'')+8 \pi_2 (3 F'^2+8) F'^2 F''-64 \pi _1 (F'^2+4) F' F''  \bigg]  , \\
 \label{eq:A2}
\mc{A}_2
&=&\frac{1}{16 \left(F'^2+8\right)^2 \left(3 F'^2+16\right)} \bigg[ -16 \pi _1 F' \big(64 \pi _2 \left(3 F'^2+8\right)
\notag \\
&& -\left(F'^2+8\right) \big(-3 F'^3\pd_1 W^1  +16 \left(F'\pd_2 W^2+\left(F'^2+2\right)\pd_2 W^1\right) +32\pd_1 W^2\big)\big)
\notag\\
&&+\left(F'^2+8\right)^2 \left(\left(F'^2-8\right)\pd_2 W^1+4 F' \left( \pd_2 W^2 \pd_2 W^2+\pd_2 W^2\right)+8\pd_1 W^2\right)^2
\notag\\
&&-16 \pi _2 F' \left(F'^2+8\right) \left(F' \left(\left(3 F'^2+40\right)\pd_1 W^2+12 F'\pd_1 W^1+3 \left(F'^2-8\right)\pd_2 W^1\right)-64\pd_2 W^2\right)
\notag\\
&&+4096 \pi _1^2 \left(F'^2+4\right)+256 \pi _2^2 \left(3 F'^4+16 F'^2+64\right)  \bigg]\,.
\ea
As soon as $\mc{A}_0 \ne 0$,  $W^0$ enters quadratically and it no longer imposes an additional first--class constraint. Rather one can easily integrate it out giving rise to the following Hamiltonian
\be
\mc{H}^{(2)}_1 = \mc{A}_2 - \frac{\mc{A}^2_1}{4\mc{A}_0}\,.
\ee

\subsection{Quadratic Lagrangian on the perturbed background}
\label{sec:le}

In this appendix we consider the Lagrangian \eqref{K2lag} and look at fluctuations $V^\ai$ living on top of the background $\bar\phi^\ai=x^\ai+\ep \bar B^\ai$,
\ba
\phi^\ai=x^\ai+\ep \bar B^\ai+\varepsilon V^\ai\,.
\ea
Since we are looking for the stability of the fluctuations $V^\ai$, it is sufficient to construct the Lagrangian and Hamiltonian at quadratic order in fluctuations, \ien, to second order in $\varepsilon$. Moreover, we treat the background $\bar \phi^\ai$ perturbatively and for the sake of this analysis it will be sufficient to work to second order in $\ep$.
To that order in perturbations, the explicit form of Lagrangian (\ref{V2ndLag}) is then given by
\bal
\mc{L} &= -  \pd_{[\mu} V_{\nu]} \pd^{\mu} V^{\nu}
\nn
&~~~
+\frac{\ep}{4} \bigg[-\pd^\ri {\bar B} ^\mu
   \pd_\mu V^\nu
   \pd_\ri V_\nu -4 \pd_\nu V^\nu
   \pd^\ri {\bar B} ^\mu
   \pd_\mu V_\ri +4 \pd_\nu V^\nu
   \pd^\ri {\bar B} ^\mu
   \pd_\ri V_\mu
      -\pd^\ri {\bar B} ^\mu  \pd^\nu V_\mu
   \pd_\ri V_\nu
   \nn
   &~~~
   -\pd^\ri {\bar B}
   ^\mu  \pd^\nu V_\mu
   \pd_\nu V_\ri +3
   \pd^\ri {\bar B} ^\mu  \pd_\mu V_\nu
   \pd^\nu V_\ri -2 \pd_\mu {\bar B} ^\mu
   \pd_\ri V_\nu
   \pd^\nu V^\ri
   +2 \pd_\mu {\bar B} ^\mu
   \pd_\nu V_\ri
   \pd^\nu V^\ri \bigg]
   \nn
   &~~~
+\frac{\ep^2}{32} \bigg[ 2 \pd_\mu V^\si
   \pd^\ri {\bar B} ^\mu
   \pd_\nu V_\si
   (\pd_\ri {\bar B} ^\nu -\pd^\nu {\bar B} _\ri) -\pd_\mu {\bar B} _\ri
   \pd^\mu V^\si
   \pd_\nu V_\si
   \pd^\nu {\bar B} ^\ri -7 \pd_\mu {\bar B} ^\mu
   \pd_\ri V^\si
   \pd_\nu V_\si
   \pd^\nu {\bar B} ^\ri
    \nn
   &~~~
   -\pd_\ri V_\si \pd^\ri {\bar B} ^\mu
   \pd_\nu {\bar B} _\mu
   \pd^\nu V^\si -\pd_\mu {\bar B} ^\mu
   \pd_\ri V_\si
   \pd_\nu {\bar B} ^\ri
   \pd^\nu V^\si -\pd_\ri V_\nu \pd^\ri {\bar B} ^\mu  \pd_\nu {\bar B} _\mu
   \pd_\si V^\si
      \nn
   &~~~
   -6
   \pd^\ri {\bar B} ^\mu
   \pd_\nu V_\ri  \pd^\nu {\bar B} _\mu
   \pd_\si V^\si +18
   \pd_\mu V_\nu  \pd^\ri {\bar B} ^\mu
   \pd^\nu {\bar B} _\ri
   \pd_\si V^\si
   -6 \pd^\ri {\bar B} ^\mu  \pd_\nu V_\mu
   \pd^\nu {\bar B} _\ri
   \pd_\si V^\si
      \nn
   &~~~
   -\pd_\mu {\bar B}
   _\ri  \pd_\nu V^\mu
   \pd^\nu {\bar B} ^\ri
   \pd_\si V^\si +2
   \pd_\mu {\bar B} _\nu  \pd^\nu {\bar B} ^\ri  (3
   \pd^\mu V_\si
   \pd_\ri V^\si -(\pd^\mu V
   _\ri -3 \pd_\ri V^\mu )
   \pd_\si V^\si )
      \nn
   &~~~
   +3
   \pd_\mu V^\nu
   \pd_\ri V_\si
   \pd^\ri {\bar B} ^\mu
   \pd^\si {\bar B} _\nu +2
   \pd_\ri V_\mu
   \pd^\ri {\bar B} ^\mu
   \pd_\si V^\nu
   \pd^\si {\bar B} _\nu +4
   \pd_\ri V_\si
   \pd^\ri {\bar B} ^\mu  \pd_\nu V_\mu
   \pd^\si {\bar B} ^\nu
      \nn
   &~~~
   -10
   \pd_\mu V_\si
   \pd^\ri {\bar B} ^\mu
   \pd_\nu V_\ri
   \pd^\si {\bar B} ^\nu +8
   \pd_\mu V_\ri
   \pd^\ri {\bar B} ^\mu
   \pd_\nu V_\si
   \pd^\si {\bar B} ^\nu
   -8
   \pd_\ri {\bar B} ^\nu
   \pd^\ri {\bar B} ^\mu  \pd_\nu V_\mu
   \pd_\si V^\si
   \nn
   &~~~
   -16 \pd_\ri V_\mu
   \pd^\ri {\bar B} ^\mu
   \pd_\nu V_\si
   \pd^\si {\bar B} ^\nu -2
   \pd_\ri V_\nu
   \pd^\ri {\bar B} ^\mu
   \pd_\si V_\mu
   \pd^\si {\bar B} ^\nu +4
   \pd^\ri {\bar B} ^\mu
   \pd_\nu V_\ri
   \pd_\si V_\mu
   \pd^\si {\bar B} ^\nu
      \nn
   &~~~
   +\pd^\ri {\bar B} ^\mu \pd_\nu V_\mu
   \pd_\si V_\ri
   \pd^\si {\bar B} ^\nu +6
   \pd_\ri V_\mu
   \pd^\ri {\bar B} ^\mu
   \pd_\si V_\nu
   \pd^\si {\bar B} ^\nu +4
   \pd_\ri {\bar B} ^\nu
   \pd^\ri {\bar B} ^\mu
   \pd_\nu V_\si
   \pd^\si V_\mu
      \nn
   &~~~
   +4
   \pd^\ri {\bar B} ^\mu
   \pd_\nu V_\si
   \pd^\nu {\bar B} _\ri
   \pd^\si V_\mu -2
   \pd_\ri {\bar B} ^\nu
   \pd^\ri {\bar B} ^\mu
   \pd_\si V_\nu
   \pd^\si V_\mu +4
   \pd^\ri {\bar B} ^\mu
   \pd^\nu {\bar B} _\ri
   \pd_\si V_\nu
   \pd^\si V_\mu
   \nn
   &~~~
   +4 \pd^\ri {\bar B} ^\mu
   \pd_\nu V_\si  \pd^\nu {\bar B} _\mu
   \pd^\si V_\ri -8
   \pd_\mu {\bar B} ^\mu  \pd_\nu V_\si
   \pd^\nu {\bar B} ^\ri
   \pd^\si V_\ri +2
   \pd^\ri {\bar B} ^\mu  \pd^\nu {\bar B} _\mu
   \pd_\si V_\nu
   \pd^\si V_\ri
      \nn
   &~~~
   -8
   \pd_\mu {\bar B} ^\mu  \pd^\nu {\bar B} ^\ri
   \pd_\si V_\nu
   \pd^\si V_\ri -20
   \pd_\mu V_\si
   \pd^\ri {\bar B} ^\mu
   \pd^\nu {\bar B} _\ri
   \pd^\si V_\nu +24 \pd_\mu {\bar B} ^\mu
   \pd_\ri V_\si
   \pd^\nu {\bar B} ^\ri
   \pd^\si V_\nu
      \nn
   &~~~
   +4
   \pd_\mu {\bar B} _\ri
   \pd^\ri {\bar B} ^\mu
   \pd_\nu V_\si
   \pd^\si V^\nu -4
   \pd_\ri {\bar B} _\mu
   \pd^\ri {\bar B} ^\mu
   \pd_\nu V_\si
   \pd^\si V^\nu
      \nn
   &~~~
   -4
   \pd_\mu {\bar B} _\ri
   \pd^\ri {\bar B} ^\mu
   \pd_\si V_\nu
   \pd^\si V^\nu +4
   \pd_\ri {\bar B} _\mu
   \pd^\ri {\bar B} ^\mu
   \pd_\si V_\nu
   \pd^\si V^\nu \bigg]
   +\mc{O}(\ep^3)\,.
\eal

\bibliographystyle{JHEPmodplain}
\bibliography{refs}

\providecommand{\href}[2]{#2}\begingroup\raggedright\begin{thebibliography}{10}

\bibitem{ArkaniHamed:2002sp}
N.~Arkani-Hamed, H.~Georgi, and M.~D. Schwartz, {\it {Effective field theory
  for massive gravitons and gravity in theory space}},  {\sl Annals Phys.} {\bf
  305} (2003) 96--118, [\href{http://arxiv.org/abs/hep-th/0210184}{{\sf
  arXiv:hep-th/0210184}}],
  [\href{http://dx.doi.org/10.1016/S0003-4916(03)00068-X}{{\sf
  doi:10.1016/S0003-4916(03)00068-X}}].

\bibitem{Boulware:1973my}
D.~Boulware and S.~Deser, {\it {Can gravitation have a finite range?}},  {\sl
  Phys.Rev.} {\bf D6} (1972) 3368--3382,
  [\href{http://dx.doi.org/10.1103/PhysRevD.6.3368}{{\sf
  doi:10.1103/PhysRevD.6.3368}}].

\bibitem{Deffayet:2005ys}
C.~Deffayet and J.-W. Rombouts, {\it {Ghosts, strong coupling and accidental
  symmetries in massive gravity}},  {\sl Phys.Rev.} {\bf D72} (2005) 044003,
  [\href{http://arxiv.org/abs/gr-qc/0505134}{{\sf arXiv:gr-qc/0505134}}],
  [\href{http://dx.doi.org/10.1103/PhysRevD.72.044003}{{\sf
  doi:10.1103/PhysRevD.72.044003}}].

\bibitem{Creminelli:2005qk}
P.~Creminelli, A.~Nicolis, M.~Papucci, and E.~Trincherini, {\it {Ghosts in
  massive gravity}},  {\sl JHEP} {\bf 0509} (2005) 003,
  [\href{http://arxiv.org/abs/hep-th/0505147}{{\sf arXiv:hep-th/0505147}}],
  [\href{http://dx.doi.org/10.1088/1126-6708/2005/09/003}{{\sf
  doi:10.1088/1126-6708/2005/09/003}}].

\bibitem{Vainshtein:1972sx}
A.~Vainshtein, {\it {To the problem of nonvanishing gravitation mass}},  {\sl
  Phys.Lett.} {\bf B39} (1972) 393--394,
  [\href{http://dx.doi.org/10.1016/0370-2693(72)90147-5}{{\sf
  doi:10.1016/0370-2693(72)90147-5}}].

\bibitem{vanDam:1970vg}
H.~van Dam and M.~J.~G. Veltman, {\it {Massive and massless Yang-Mills and
  gravitational fields}},  {\sl Nucl. Phys.} {\bf B22} (1970) 397--411,
  [\href{http://dx.doi.org/10.1016/0550-3213(70)90416-5}{{\sf
  doi:10.1016/0550-3213(70)90416-5}}].

\bibitem{Zakharov:1970cc}
V.~I. Zakharov, {\it {Linearized gravitation theory and the graviton mass}},
  {\sl JETP Lett.} {\bf 12} (1970) 312. [Pisma Zh. Eksp. Teor.
  Fiz.12,447(1970)].

\bibitem{deRham:2010ik}
C.~de~Rham and G.~Gabadadze, {\it {Generalization of the Fierz-Pauli Action}},
  {\sl Phys.Rev.} {\bf D82} (2010) 044020,
  [\href{http://arxiv.org/abs/1007.0443}{{\sf arXiv:1007.0443}}],
  [\href{http://dx.doi.org/10.1103/PhysRevD.82.044020}{{\sf
  doi:10.1103/PhysRevD.82.044020}}].

\bibitem{deRham:2010kj}
C.~de~Rham, G.~Gabadadze, and A.~J. Tolley, {\it {Resummation of Massive
  Gravity}},  {\sl Phys.Rev.Lett.} {\bf 106} (2011) 231101,
  [\href{http://arxiv.org/abs/1011.1232}{{\sf arXiv:1011.1232}}],
  [\href{http://dx.doi.org/10.1103/PhysRevLett.106.231101}{{\sf
  doi:10.1103/PhysRevLett.106.231101}}].

\bibitem{Hassan:2011hr}
S.~Hassan and R.~A. Rosen, {\it {Resolving the Ghost Problem in non-Linear
  Massive Gravity}},  {\sl Phys.Rev.Lett.} {\bf 108} (2012) 041101,
  [\href{http://arxiv.org/abs/1106.3344}{{\sf arXiv:1106.3344}}],
  [\href{http://dx.doi.org/10.1103/PhysRevLett.108.041101}{{\sf
  doi:10.1103/PhysRevLett.108.041101}}].

\bibitem{Hassan:2011ea}
S.~Hassan and R.~A. Rosen, {\it {Confirmation of the Secondary Constraint and
  Absence of Ghost in Massive Gravity and Bimetric Gravity}},  {\sl JHEP} {\bf
  1204} (2012) 123, [\href{http://arxiv.org/abs/1111.2070}{{\sf
  arXiv:1111.2070}}], [\href{http://dx.doi.org/10.1007/JHEP04(2012)123}{{\sf
  doi:10.1007/JHEP04(2012)123}}].

\bibitem{Deffayet:2001uk}
C.~Deffayet, G.~R. Dvali, G.~Gabadadze, and A.~I. Vainshtein, {\it
  {Nonperturbative continuity in graviton mass versus perturbative
  discontinuity}},  {\sl Phys. Rev.} {\bf D65} (2002) 044026,
  [\href{http://arxiv.org/abs/hep-th/0106001}{{\sf arXiv:hep-th/0106001}}],
  [\href{http://dx.doi.org/10.1103/PhysRevD.65.044026}{{\sf
  doi:10.1103/PhysRevD.65.044026}}].

\bibitem{deRham:2014zqa}
C.~de~Rham, {\it {Massive Gravity}},  {\sl Living Rev.Rel.} {\bf 17} (2014) 7,
  [\href{http://arxiv.org/abs/1401.4173}{{\sf arXiv:1401.4173}}],
  [\href{http://dx.doi.org/10.12942/lrr-2014-7}{{\sf
  doi:10.12942/lrr-2014-7}}].

\bibitem{Babichev:2013usa}
E.~Babichev and C.~Deffayet, {\it {An introduction to the Vainshtein
  mechanism}},  {\sl Class. Quant. Grav.} {\bf 30} (2013) 184001,
  [\href{http://arxiv.org/abs/1304.7240}{{\sf arXiv:1304.7240}}],
  [\href{http://dx.doi.org/10.1088/0264-9381/30/18/184001}{{\sf
  doi:10.1088/0264-9381/30/18/184001}}].

\bibitem{deRham:2015ijs}
C.~de~Rham, A.~J. Tolley, and S.-Y. Zhou, {\it {Non-compact nonlinear sigma
  models}},  \href{http://arxiv.org/abs/1512.06838}{{\sf arXiv:1512.06838}}.

\bibitem{deRham:2011rn}
C.~de~Rham, G.~Gabadadze, and A.~J. Tolley, {\it {Ghost free Massive Gravity in
  the St\"uckelberg language}},  {\sl Phys.Lett.} {\bf B711} (2012) 190--195,
  [\href{http://arxiv.org/abs/1107.3820}{{\sf arXiv:1107.3820}}],
  [\href{http://dx.doi.org/10.1016/j.physletb.2012.03.081}{{\sf
  doi:10.1016/j.physletb.2012.03.081}}].

\bibitem{Kogan:2000uy}
I.~I. Kogan, S.~Mouslopoulos, and A.~Papazoglou, {\it {The m ---> 0 limit for
  massive graviton in dS(4) and AdS(4): How to circumvent the van
  Dam-Veltman-Zakharov discontinuity}},  {\sl Phys. Lett.} {\bf B503} (2001)
  173--180, [\href{http://arxiv.org/abs/hep-th/0011138}{{\sf
  arXiv:hep-th/0011138}}],
  [\href{http://dx.doi.org/10.1016/S0370-2693(01)00209-X}{{\sf
  doi:10.1016/S0370-2693(01)00209-X}}].

\bibitem{Porrati:2000cp}
M.~Porrati, {\it {No van Dam-Veltman-Zakharov discontinuity in AdS space}},
  {\sl Phys. Lett.} {\bf B498} (2001) 92--96,
  [\href{http://arxiv.org/abs/hep-th/0011152}{{\sf arXiv:hep-th/0011152}}],
  [\href{http://dx.doi.org/10.1016/S0370-2693(00)01380-0}{{\sf
  doi:10.1016/S0370-2693(00)01380-0}}].

\bibitem{Karch:2001jb}
A.~Karch, E.~Katz, and L.~Randall, {\it {Absence of a VVDZ discontinuity in
  AdS(AdS)}},  {\sl JHEP} {\bf 12} (2001) 016,
  [\href{http://arxiv.org/abs/hep-th/0106261}{{\sf arXiv:hep-th/0106261}}],
  [\href{http://dx.doi.org/10.1088/1126-6708/2001/12/016}{{\sf
  doi:10.1088/1126-6708/2001/12/016}}].

\bibitem{Porrati:2003sa}
M.~Porrati, {\it {Higgs phenomenon for the graviton in ADS space}},  {\sl Mod.
  Phys. Lett.} {\bf A18} (2003) 1793--1802,
  [\href{http://arxiv.org/abs/hep-th/0306253}{{\sf arXiv:hep-th/0306253}}],
  [\href{http://dx.doi.org/10.1142/S0217732303011745}{{\sf
  doi:10.1142/S0217732303011745}}].

\bibitem{Porrati:2004mz}
M.~Porrati, {\it {Massive gravity in AdS and Minkowski backgrounds}},  in {\em
  {Deserfest: A celebration of the life and works of Stanley Deser.
  Proceedings, Meeting, Ann Arbor, USA, April 3-5, 2004}}, pp.~233--241, 2004.
\newblock \href{http://arxiv.org/abs/hep-th/0409172}{{\sf
  arXiv:hep-th/0409172}}.

\bibitem{deRham:2012ew}
C.~de~Rham, G.~Gabadadze, L.~Heisenberg, and D.~Pirtskhalava, {\it
  {Non-Renormalization and Naturalness in a Class of Scalar-Tensor Theories}},
  {\sl Phys.Rev.} {\bf D87} (2013) 085017,
  [\href{http://arxiv.org/abs/1212.4128}{{\sf arXiv:1212.4128}}],
  [\href{http://dx.doi.org/10.1103/PhysRevD.87.085017}{{\sf
  doi:10.1103/PhysRevD.87.085017}}].

\bibitem{deRham:2010tw}
C.~de~Rham, G.~Gabadadze, L.~Heisenberg, and D.~Pirtskhalava, {\it {Cosmic
  Acceleration and the Helicity-0 Graviton}},  {\sl Phys. Rev.} {\bf D83}
  (2011) 103516, [\href{http://arxiv.org/abs/1010.1780}{{\sf
  arXiv:1010.1780}}], [\href{http://dx.doi.org/10.1103/PhysRevD.83.103516}{{\sf
  doi:10.1103/PhysRevD.83.103516}}].

\bibitem{Aoki:2015xqa}
K.~Aoki, K.-i. Maeda, and R.~Namba, {\it {Stability of the Early Universe in
  Bigravity Theory}},  {\sl Phys. Rev.} {\bf D92} (2015), no.~4 044054,
  [\href{http://arxiv.org/abs/1506.04543}{{\sf arXiv:1506.04543}}],
  [\href{http://dx.doi.org/10.1103/PhysRevD.92.044054}{{\sf
  doi:10.1103/PhysRevD.92.044054}}].

\bibitem{Comelli:2014xga}
D.~Comelli, F.~Nesti, and L.~Pilo, {\it {Nonderivative Modified Gravity: a
  Classification}},  {\sl JCAP} {\bf 1411} (2014), no.~11 018,
  [\href{http://arxiv.org/abs/1407.4991}{{\sf arXiv:1407.4991}}],
  [\href{http://dx.doi.org/10.1088/1475-7516/2014/11/018}{{\sf
  doi:10.1088/1475-7516/2014/11/018}}].

\bibitem{DeFelice:2015hla}
A.~De~Felice and S.~Mukohyama, {\it {Minimal theory of massive gravity}},  {\sl
  Phys. Lett.} {\bf B752} (2016) 302--305,
  [\href{http://arxiv.org/abs/1506.01594}{{\sf arXiv:1506.01594}}],
  [\href{http://dx.doi.org/10.1016/j.physletb.2015.11.050}{{\sf
  doi:10.1016/j.physletb.2015.11.050}}].

\bibitem{deRham:2010gu}
C.~de~Rham and G.~Gabadadze, {\it {Selftuned Massive Spin-2}},  {\sl Phys.
  Lett.} {\bf B693} (2010) 334--338,
  [\href{http://arxiv.org/abs/1006.4367}{{\sf arXiv:1006.4367}}],
  [\href{http://dx.doi.org/10.1016/j.physletb.2010.08.043}{{\sf
  doi:10.1016/j.physletb.2010.08.043}}].

\bibitem{nsmrev}
S.~V.~Ketov, {\it {Scholarpedia}},  {\sl Scholarpedia} {\bf 4(1)} (2009) 8508,
  [\href{http://dx.doi.org/10.4249/scholarpedia.8508}{{\sf
  doi:10.4249/scholarpedia.8508}}].

\bibitem{nsmrev2}
S.~V.~Ketov, {\em {Quantum Non-linear Sigma-Models}}.
\newblock Springer-Verlag Berlin Heidelberg, 2000.

\bibitem{Zakrzewski:1989na}
W.~J. Zakrzewski, {\it {Low Dimensional Sigma Models}},  in {\em {Bristol, UK:
  Hilger (1989) 289p, In Storrs 1988, Proceedings, 4th Meeting of the Division
  of Particles and Fields of the APS* 790-793.}}, 1989.

\bibitem{Brink:1976sc}
L.~Brink, P.~Di~Vecchia, and P.~S. Howe, {\it {A Locally Supersymmetric and
  Reparametrization Invariant Action for the Spinning String}},  {\sl Phys.
  Lett.} {\bf B65} (1976) 471--474,
  [\href{http://dx.doi.org/10.1016/0370-2693(76)90445-7}{{\sf
  doi:10.1016/0370-2693(76)90445-7}}].

\bibitem{Deser:1976rb}
S.~Deser and B.~Zumino, {\it {A Complete Action for the Spinning String}},
  {\sl Phys. Lett.} {\bf B65} (1976) 369--373,
  [\href{http://dx.doi.org/10.1016/0370-2693(76)90245-8}{{\sf
  doi:10.1016/0370-2693(76)90245-8}}].

\bibitem{Polyakov:1975rr}
A.~M. Polyakov, {\it {Interaction of Goldstone Particles in Two-Dimensions.
  Applications to Ferromagnets and Massive Yang-Mills Fields}},  {\sl Phys.
  Lett.} {\bf B59} (1975) 79--81,
  [\href{http://dx.doi.org/10.1016/0370-2693(75)90161-6}{{\sf
  doi:10.1016/0370-2693(75)90161-6}}].

\bibitem{Nambu:1986ze}
Y.~Nambu, {\it {Duality and hadrodynamics, Notes prepared for the Copenhagen
  High Energy Symposium (1970) [Reprinted: Broken Symmetry: Selected Papers of
  Y. Nambu, eds. T.~Eguchi and K.~Nishijima, Singapore: World Scientific,
  1995]}}, .

\bibitem{Goto:1971ce}
T.~Goto, {\it {Relativistic quantum mechanics of one-dimensional mechanical
  continuum and subsidiary condition of dual resonance model}},  {\sl Prog.
  Theor. Phys.} {\bf 46} (1971) 1560--1569,
  [\href{http://dx.doi.org/10.1143/PTP.46.1560}{{\sf
  doi:10.1143/PTP.46.1560}}].

\bibitem{Hara:1971ur}
O.~Hara, {\it {On origin and physical meaning of ward-like identity in
  dual-resonance model}},  {\sl Prog. Theor. Phys.} {\bf 46} (1971) 1549--1559,
  [\href{http://dx.doi.org/10.1143/PTP.46.1549}{{\sf
  doi:10.1143/PTP.46.1549}}].

\bibitem{Becker:2007zj}
K.~Becker, M.~Becker, and J.~H. Schwarz, {\em {String theory and M-theory: A
  modern introduction}}.
\newblock Cambridge University Press, 2006.

\bibitem{Cremmer:1979up}
E.~Cremmer and B.~Julia, {\it {The SO(8) Supergravity}},  {\sl Nucl. Phys.}
  {\bf B159} (1979) 141,
  [\href{http://dx.doi.org/10.1016/0550-3213(79)90331-6}{{\sf
  doi:10.1016/0550-3213(79)90331-6}}].

\bibitem{VanNieuwenhuizen:1981ae}
P.~Van~Nieuwenhuizen, {\it {Supergravity}},  {\sl Phys. Rept.} {\bf 68} (1981)
  189--398, [\href{http://dx.doi.org/10.1016/0370-1573(81)90157-5}{{\sf
  doi:10.1016/0370-1573(81)90157-5}}].

\bibitem{deRham:2014lqa}
C.~De~Rham, L.~Keltner, and A.~J. Tolley, {\it {Generalized galileon duality}},
   {\sl Phys. Rev.} {\bf D90} (2014), no.~2 024050,
  [\href{http://arxiv.org/abs/1403.3690}{{\sf arXiv:1403.3690}}],
  [\href{http://dx.doi.org/10.1103/PhysRevD.90.024050}{{\sf
  doi:10.1103/PhysRevD.90.024050}}].

\bibitem{GellMann:1960np}
M.~Gell-Mann and M.~Levy, {\it {The axial vector current in beta decay}},  {\sl
  Nuovo Cim.} {\bf 16} (1960) 705,
  [\href{http://dx.doi.org/10.1007/BF02859738}{{\sf doi:10.1007/BF02859738}}].

\bibitem{Arnowitt:1962hi}
R.~L. Arnowitt, S.~Deser, and C.~W. Misner, {\it {The Dynamics of general
  relativity}},  {\sl Gen.Rel.Grav.} {\bf 40} (2008) 1997--2027,
  [\href{http://arxiv.org/abs/gr-qc/0405109}{{\sf arXiv:gr-qc/0405109}}],
  [\href{http://dx.doi.org/10.1007/s10714-008-0661-1}{{\sf
  doi:10.1007/s10714-008-0661-1}}].

\bibitem{Berezhiani:2013dw}
L.~Berezhiani, G.~Chkareuli, and G.~Gabadadze, {\it {Restricted Galileons}},
  {\sl Phys. Rev.} {\bf D88} (2013) 124020,
  [\href{http://arxiv.org/abs/1302.0549}{{\sf arXiv:1302.0549}}],
  [\href{http://dx.doi.org/10.1103/PhysRevD.88.124020}{{\sf
  doi:10.1103/PhysRevD.88.124020}}].

\bibitem{Berezhiani:2013dca}
L.~Berezhiani, G.~Chkareuli, C.~de~Rham, G.~Gabadadze, and A.~J. Tolley, {\it
  {Mixed Galileons and Spherically Symmetric Solutions}},  {\sl Class. Quant.
  Grav.} {\bf 30} (2013) 184003, [\href{http://arxiv.org/abs/1305.0271}{{\sf
  arXiv:1305.0271}}],
  [\href{http://dx.doi.org/10.1088/0264-9381/30/18/184003}{{\sf
  doi:10.1088/0264-9381/30/18/184003}}].

\bibitem{Alberte:2014bua}
L.~Alberte and A.~Khmelnitsky, {\it {Stability of Massive Gravity Solutions for
  Holographic Conductivity}},  {\sl Phys. Rev.} {\bf D91} (2015), no.~4 046006,
  [\href{http://arxiv.org/abs/1411.3027}{{\sf arXiv:1411.3027}}],
  [\href{http://dx.doi.org/10.1103/PhysRevD.91.046006}{{\sf
  doi:10.1103/PhysRevD.91.046006}}].

\bibitem{Alberte:2013sma}
L.~Alberte and A.~Khmelnitsky, {\it {Reduced massive gravity with two
  Stückelberg fields}},  {\sl Phys. Rev.} {\bf D88} (2013), no.~6 064053,
  [\href{http://arxiv.org/abs/1303.4958}{{\sf arXiv:1303.4958}}],
  [\href{http://dx.doi.org/10.1103/PhysRevD.88.064053}{{\sf
  doi:10.1103/PhysRevD.88.064053}}].

\bibitem{Hassan:2011tf}
S.~F. Hassan, R.~A. Rosen, and A.~Schmidt-May, {\it {Ghost-free Massive Gravity
  with a General Reference Metric}},  {\sl JHEP} {\bf 02} (2012) 026,
  [\href{http://arxiv.org/abs/1109.3230}{{\sf arXiv:1109.3230}}],
  [\href{http://dx.doi.org/10.1007/JHEP02(2012)026}{{\sf
  doi:10.1007/JHEP02(2012)026}}].

\bibitem{DeFelice:2012mx}
A.~De~Felice, A.~E. Gumrukcuoglu, and S.~Mukohyama, {\it {Massive gravity:
  nonlinear instability of the homogeneous and isotropic universe}},  {\sl
  Phys. Rev. Lett.} {\bf 109} (2012) 171101,
  [\href{http://arxiv.org/abs/1206.2080}{{\sf arXiv:1206.2080}}],
  [\href{http://dx.doi.org/10.1103/PhysRevLett.109.171101}{{\sf
  doi:10.1103/PhysRevLett.109.171101}}].

\bibitem{Hinterbichler:2012cn}
K.~Hinterbichler and R.~A. Rosen, {\it {Interacting Spin-2 Fields}},  {\sl
  JHEP} {\bf 1207} (2012) 047, [\href{http://arxiv.org/abs/1203.5783}{{\sf
  arXiv:1203.5783}}], [\href{http://dx.doi.org/10.1007/JHEP07(2012)047}{{\sf
  doi:10.1007/JHEP07(2012)047}}].

\bibitem{Golovnev:2011aa}
A.~Golovnev, {\it {On the Hamiltonian analysis of non-linear massive gravity}},
   {\sl Phys. Lett.} {\bf B707} (2012) 404--408,
  [\href{http://arxiv.org/abs/1112.2134}{{\sf arXiv:1112.2134}}],
  [\href{http://dx.doi.org/10.1016/j.physletb.2011.12.064}{{\sf
  doi:10.1016/j.physletb.2011.12.064}}].

\bibitem{Hassan:2012qv}
S.~Hassan, A.~Schmidt-May, and M.~von Strauss, {\it {Proof of Consistency of
  Nonlinear Massive Gravity in the St\'uckelberg Formulation}},  {\sl
  Phys.Lett.} {\bf B715} (2012) 335--339,
  [\href{http://arxiv.org/abs/1203.5283}{{\sf arXiv:1203.5283}}],
  [\href{http://dx.doi.org/10.1016/j.physletb.2012.07.018}{{\sf
  doi:10.1016/j.physletb.2012.07.018}}].

\end{thebibliography}\endgroup

\end{document}